\documentclass{aa}
\usepackage{bibtex/natbib}
\usepackage{xspace}
\usepackage{amssymb}
\usepackage{amsmath}
\usepackage{graphicx}
\usepackage{verbatim} 
\usepackage{fancyhdr}

\begin{document}

\title{The first stages of planet formation in binary systems:\\
How far can dust coagulation proceed? }
\authorrunning{A. Zsom et al.}
\titlerunning{Planet formation in binary systems}
\author{A. Zsom, Zs. S\'andor, C. P. Dullemond}
\institute{Max Planck Research Group at the Max-Planck-Institute f\"ur Astronomie, K\"onigstuhl 17, D-69117 Heidelberg, Germany. Email: \texttt{zsom@mpia.de}}
\date{\today}

 \abstract
   {Based on current theoretical models, giant planets can form via gravitational instability or core accretion. The terrestrial planets are ``failed cores'' in the core accretion paradigm that do not evolve into gas giants. Planets residing in binary systems put a strong constraint on any planet formation theory as the model must work in binary systems too, in spite of the strong perturbations from the secondary star.}
   {We examine the first phase of the core accretion model, namely the dust growth/fragmentation in binary systems. In our model, a gas and dust disk is present around the primary star and is perturbed by the secondary. We study the effects of a secondary with/without eccentricity on the dust population to determine what sizes the aggregates can reach and how that compares to the dust population in disks around single stars.}
   {The equation of motion of dust aggregates including gas drag and gravitational forces from both stars, as well as the hydrodynamical equations for the gas disk are solved. We determine the velocity of these particles relative to the gas (or relative to micron sized, well coupled dust particles) and we construct a collision model with growth and erosion. Particles below the critical fragmentation velocity increase in mass by sweeping up the small dust, particles above this critical speed are eroded due to collisions with the small dust population. We use this model to determine the sizes that aggregates can reach in the disk and we also determine the equilibrium mass and stopping time distributions between growth and erosion.}
  {We find that the secondary star has two effects on the dust population. 1.) The disk is truncated due to the presence of the secondary star and the maximum mass of the particles is decreased in the lowered gas densities. This effect is dominant in the outer disk. 2.) The perturbation of the secondary pumps up the eccentricity of the gas disk, which in turn increases the relative velocity between the dust and the gas. Therefore the maximum particle sizes are further decreased. The second effect of the secondary influences the entire disk. Coagulation is efficiently reduced even at the very inner parts of the disk. The average mass of the particles is reduced by four orders of magnitude (as a consequence, the stopping time is reduced by one order of magnitude) in disks around binary systems compared to dust in disks around single stars.}
 {}

\maketitle

\section{Introduction}
Among all the planets known around main sequence stars, roughly 10\% resides in binary systems. All of these systems have a so called S-type configuration \citep{Dvorak1986} where the planet orbits around one of the stars and the companion acts as a perturber. Most of these binaries have separations larger than 100--300 AU in which case the secondaries play a more limited role in the formation, dynamical evolution and migration of the planets \citep{Desidera2007, Duchene2010}. However, there are some systems with small separations ($\sim$ 20 AU). These are $\gamma$ Cep with a separation of 18.5 AU \citep{Hatzes2003, Neuhauser2007}, GL 83 with a separation of 18.4 AU \citep{Lagrange2006}, HD 41004 with a separation of 23 AU \citep{Zucker2003} and HD 196885 with a separation of 17 AU \citep{Correia2008}. The mere existence of these systems puts a very strong constraint on any planet formation theory as the models have to be able to produce planets in these close binaries as well.

The core accretion paradigm of planet formation \citep{Mizuno1980, Pollack1996} has three stages, starting with the coagulation of sub-micron sized grains which leads to the formation of planetesimals \citep{Safronov1969, Weidenschilling1993, Blum2008}. The next stage is the formation of the protoplanetary cores from the planetesimals \citep{Weidenschilling1980, Nakagawa1983, Wetherill:1990p85, Tanaka2005}. Finally, the last stage of planet formation is gas accretion onto these cores, or - in case the gas is absent - gravitational encounters between these protoplanets, which result in a chaotic impact phase, until orbital stability is achieved \citep{Chambers2001, Kokubo2006, Thommes2008}. 

This picture is already very complex and has many unresolved problems, we only mention here a few. The dust coagulation process, the first stage of planet formation has to overcome the charging barrier \citep{Okuzumi2009}, bouncing barrier \citep{Zsom2010}, radial drift barrier \citep{Weidenschilling1977} and the fragmentation barrier \citep{Blum2008}. Due to these uncertainties, the initial planetesimal size distribution is poorly constrained \citep{Morbidelli2009}, therefore the second and third stages are also affected by these barriers. The presence of a close companion further complicates this processes in many ways. 

The tidal torques of the companion generate strong spiral arms in the disk around the primary. Angular momentum is transferred to the binary orbit which will truncate and restructure the disk \citep{Artymowicz1994, Armitage1999, Kley2008a}. This dynamical effect of the secondary has several consequences which might influence planet formation: it decreases the lifetime of the disk, increases the temperature of the disk and modifies the stabile orbits around the primary. 

The evolution of planetesimals is also influenced in a binary system. The perturbation of the secondary increases the relative velocity of the planetesimals and/or creates unstable regions where the planetary building blocks cannot maintain a stable orbit as shown by e.g., \cite{Heppenheimer1978}, \cite{Whitmire1998}, \cite{Th'ebault2004}. The increased relative velocities can then lead to the disruption of the planetesimals. However, \cite{Marzari2000} showed that the combined effects of the gravitational perturbation and the gas drag may increase the efficiency of their accretion by reducing their relative velocity and produce, later on, terrestrial planets. \cite{Marzari2009} calculated the relative velocity of planetesimals in highly inclined systems and concluded that planet formation appears possible for inclinations as high as 10$^{\circ}$, if the separation between the stars is larger than 70 AU. The region where planetesimals can accumulate into protoplanets shrinks consistently for lower binary separations.

Numerical simulations have shown that it might be possible to form giant planets via gravitational instability which is the other planet formation paradigm (\cite{Boss2006}, \cite{Mayer2007} and references therein), although some of these studies had to assume larger than 20 AU separations (50-100 AU), or rapid cooling of the gas. We, however, concentrate on the initial stages of the core accretion model, thus we study the formation of terrestrial planets and/or giant planets formed by core accretion. 

Several authors studied planet formation and migration in the $\gamma$ Cephei system \citep{Dvorak2004, Th'ebault2004, Haghighipour2006, Verrier2006, Kley2008, Jang-Condell2008, Paardekooper2008, Paardekooper2010}. We also consider the parameters of this binary (values taken from \cite{Torres2007}) as a starting point in our study. We focus on the motion and relative velocity of dust particles, thus investigate the first stage of planet formation in this environment. First, we produce a surface density profile of the gas, which is in quasi-equilibrium and truncated by the secondary. Once the equilibrium configuration is reached, we follow the motion of $10^5$ large dust particles. The small, sub-micron sized particles are well coupled to the gas, therefore their relative velocity compared to the gas is zero. The relative velocity of larger particles (pebbles, boulders) can be therefore considered as the collision speed between a well coupled sub-micron sized particle and the large particle in question. Based on this recognition, we construct a simple collision model in which the larger particles lose mass, if their relative velocity is larger than the critical fragmentation velocity, and increase mass below this critical speed. The critical fragmentation velocity can be determined by laboratory experiments and its value is usually 1 m/s for silicates \citep{Blum2008}. This way we determine the maximum particle size reachable in these disk configurations. We examine the properties of these particles to find out how far can coagulation proceed in such an environment, what is the size of the building blocks of planetesimals in a binary environment and how does this compare to the sizes reachable in disks around single stars.

The paper is organized as follows: in Sect. 2 we describe the model set-up of the gas disks and the numerical method to integrate the particle motion and mass; in Sect. 3 we present our results. Finally, we discuss the results and draw conclusions in Sects. 4 and 5. 

\begin{figure*}
\includegraphics[width=0.5\textwidth]{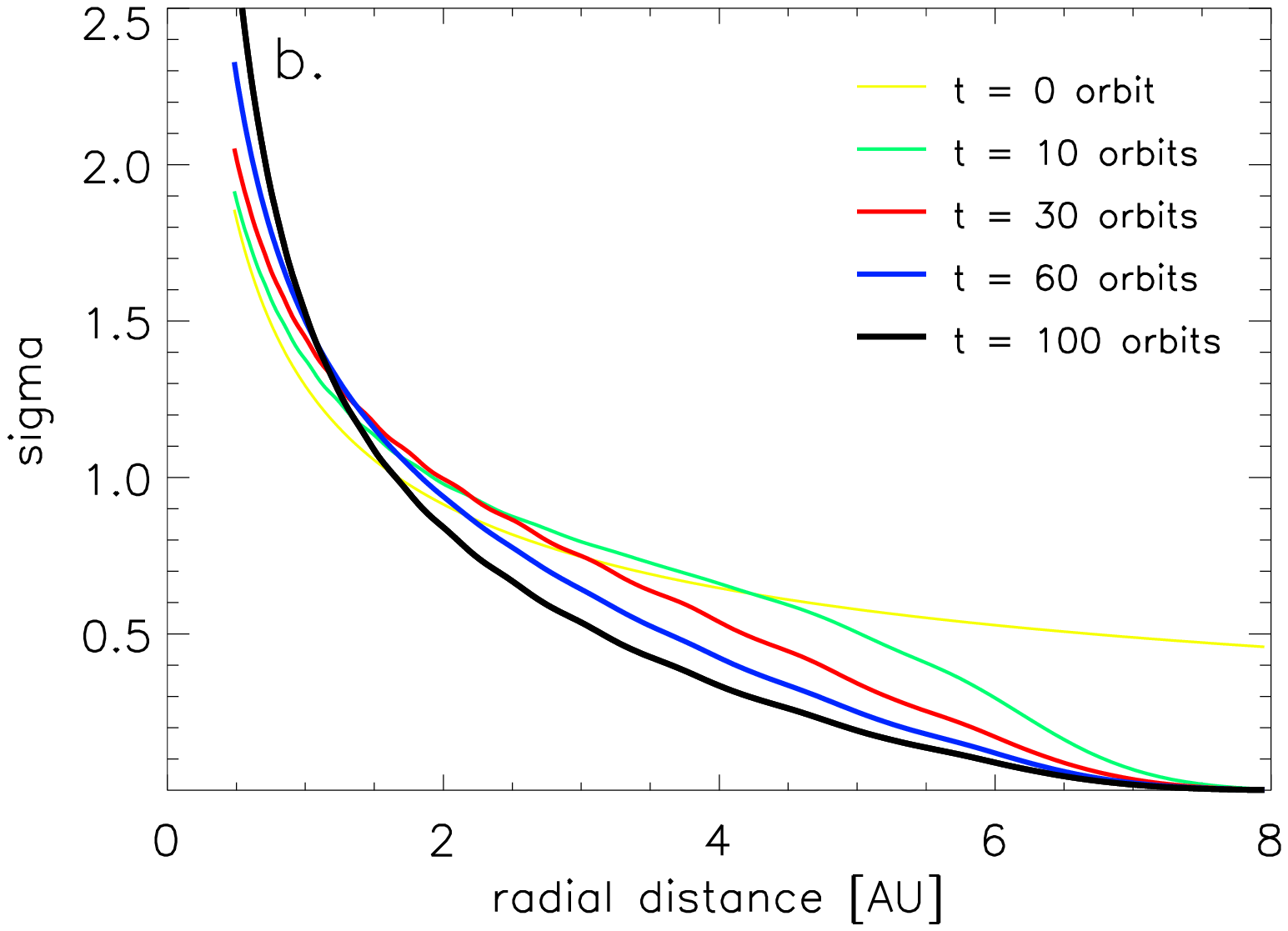}
\includegraphics[width=0.5\textwidth]{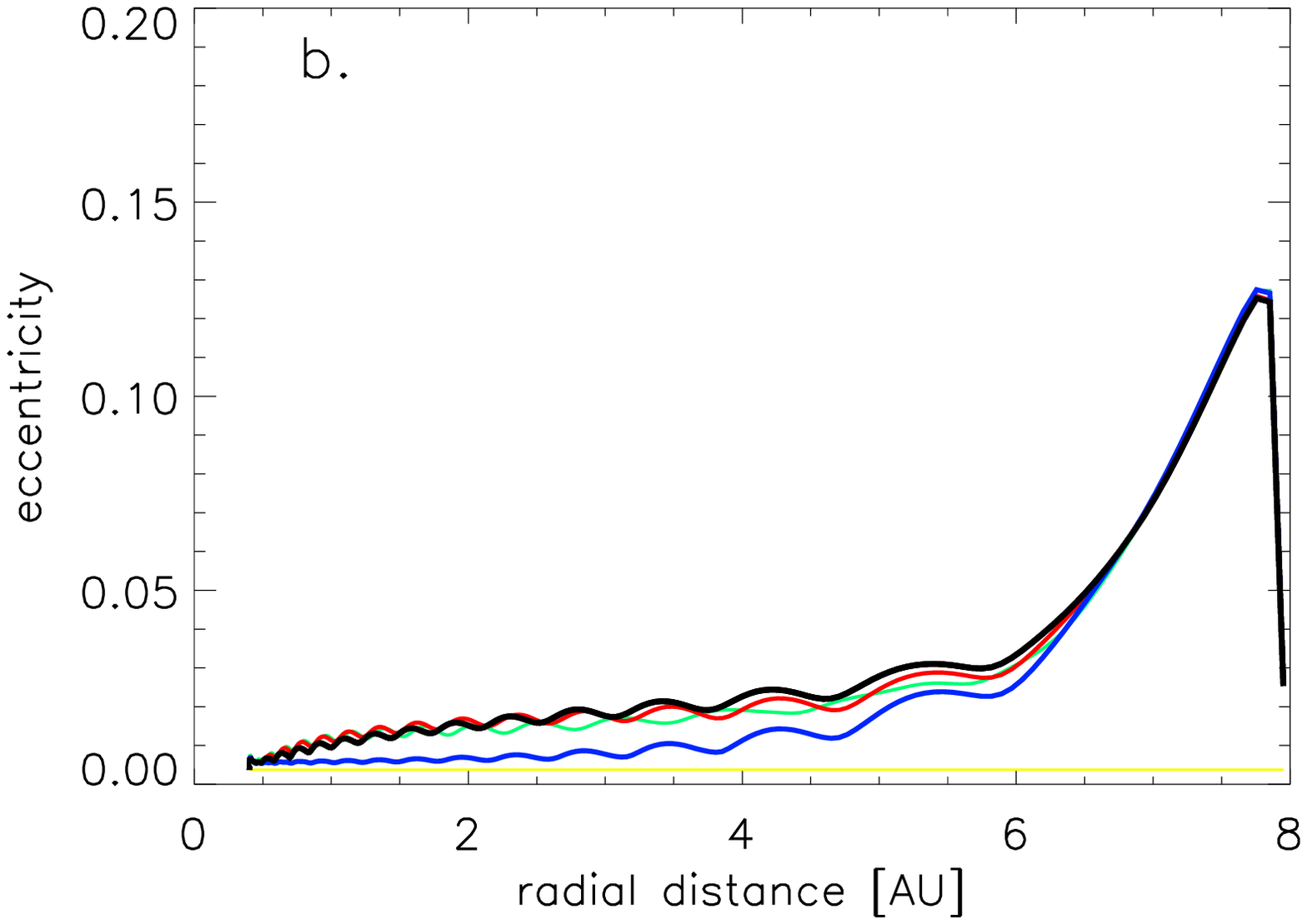}
\caption{The radial profiles of the azimuthally averaged surface density (a.) and eccentricity (b.) in the presence of a secondary on a circular orbit (`no\_ecc' set up).}
\label{fig:sigma0}
\end{figure*}

\begin{figure*}
\includegraphics[width=0.5\textwidth]{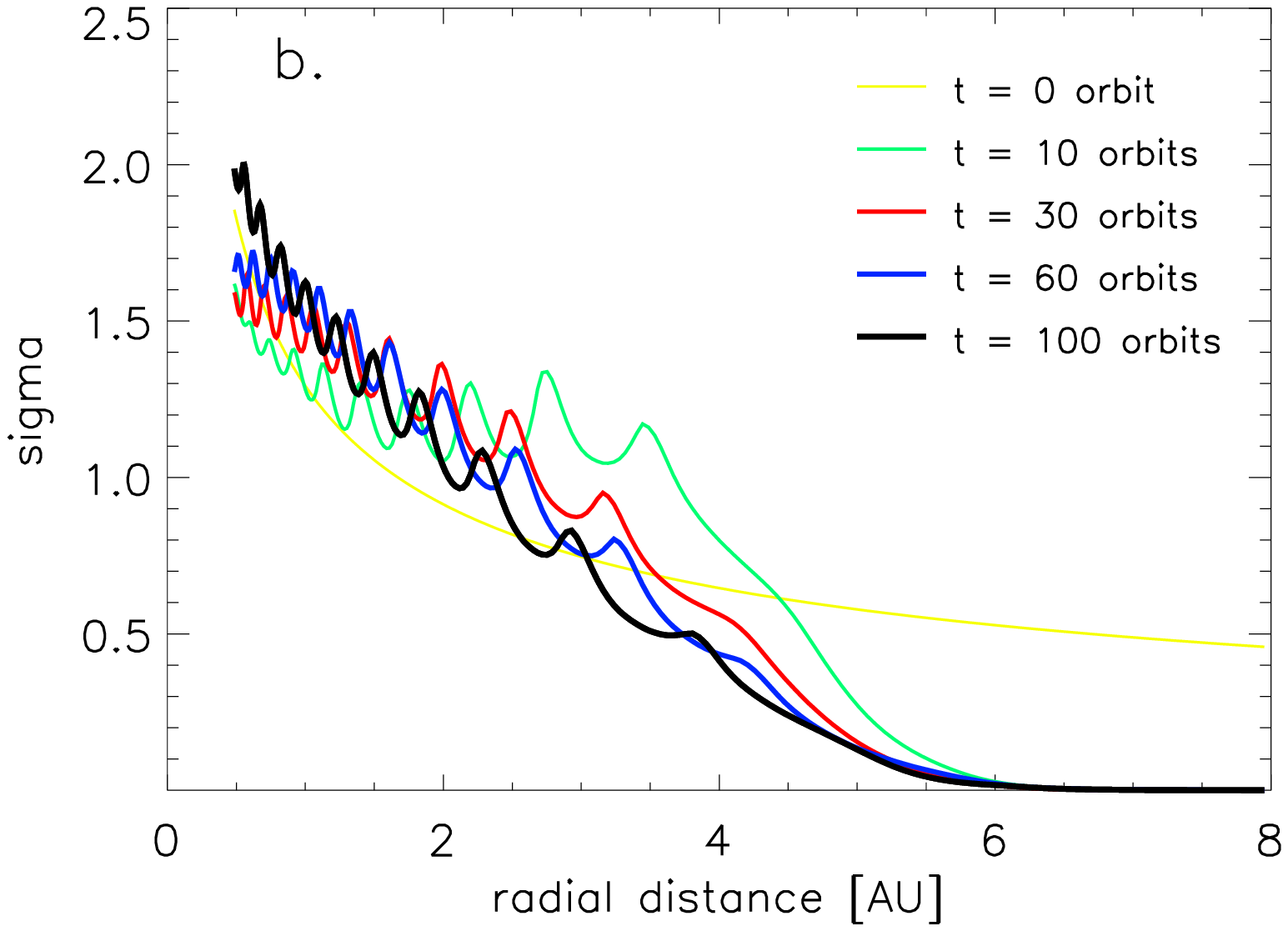}
\includegraphics[width=0.5\textwidth]{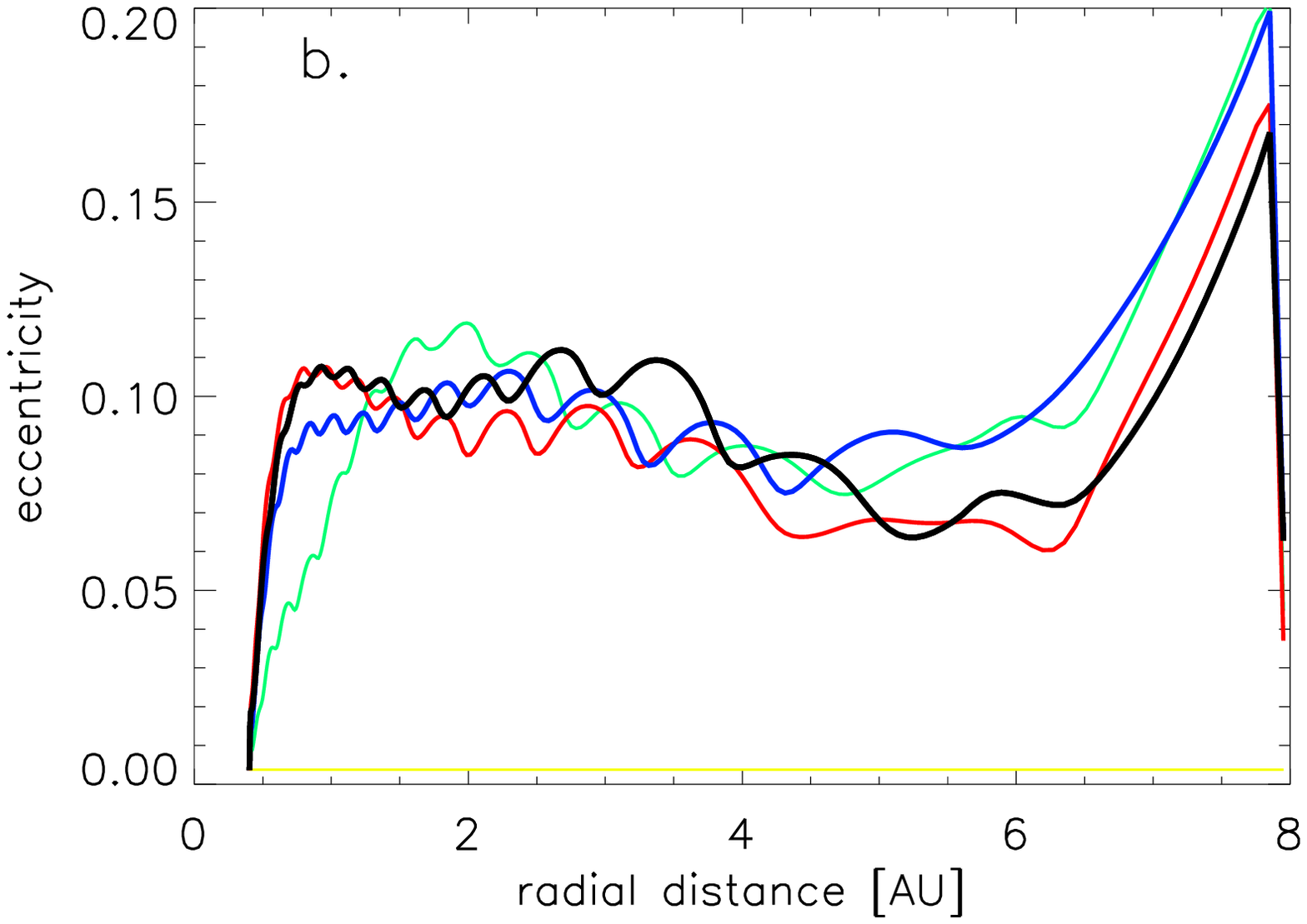}
\caption{The radial profiles of the azimuthally averaged surface density (a.) and eccentricity (b.) in the presence of a secondary for the `fiducial' simulation.}
\label{fig:sigma1}
\end{figure*}

\section{Numerical modeling}
\subsection{Gas}
We assume that the primary star is surrounded by a flat, non-flaring and non-selfgravitating disk which is coplanar with the orbit of the secondary. Therefore, we perform 2D hydrodynamical simulations using the FARGO code \citep{Masset:2000p72}, which solves the vertically integrated hydrodynamical equations. This approximation is valid if the pressure scale height of the disk ($H_p$) is small compared to the radial distance from the star ($h=H_p/r \ll 1$, where $h$ is the aspect ratio of the disk). The intrinsic turbulence of the gas is assumed to be of $\alpha$-type \citep{Shakura1973}, the turbulent kinematic viscosity ($\nu$) is
\begin{equation}
\label{eq:nuT}
\nu = \alpha c_s H_p,
\end{equation}
where $c_s$ is the isothermal sound speed. The value of the $\alpha$ parameter reflects the strengths of the turbulence in the disk, typical values are between $10^{-5} - 10^{-2}$. Finally, we assume that the disk is locally isothermal, the sound speed is
\begin{equation}
c_s = v_{\mathrm{Kep}} h,
\end{equation}
where $v_{\mathrm{Kep}}$ is the Kepler velocity. FARGO uses spherical coordinates ($r$, $\varphi$) and the origin of the coordinate system lies in the center of the primary star at $r=0$, $\varphi=0$. The coordinate system is non-rotating.

We perform 8 simulations using different initial conditions. Our `fiducial' simulation uses the stellar parameters of the $\gamma$ Cep system, namely we choose the mass ratio between the secondary and the primary to be 0.286 ($M_1 = 1.4 M_{\sun}$, $M_2 = 0.4 M_{\sun}$), the eccentricity of the secondary to be 0.4, and their separation $a_{\mathrm{bin}} = 20$ AU, which translates into an orbital period of $P = 66.7$ yrs \citep{Neuhauser2007}. We list the parameters of all the simulations in Table \ref{table:ini}. We perform a simulation without a secondary (`single') assuming a simple power-law disk profile and determine the maximum particle size in this disk. Comparing the outcome of the `single' simulation to the `no\_ecc' simulation, in which the companion has zero eccentricity, enables us to determine the effects of the presence of a secondary on the dust population. Similarly, comparing the `no\_ecc' results to the outcome of the `fiducial' simulation makes it possible to determine the effects of an eccentric companion on the dust population, etc. For the rest of the simulations, the initial parameters are chosen such that only one parameter has a different value than our `fiducial' simulation except for the `alpha' simulation. We found that with a non-reflecting boundary condition with $\alpha=10^{-3}$, the eccentricity of the gas is not converging within 200 secondary orbits. We found that the average eccentricity of the gas disk is 0.003, 0.07, 0.17, 0.23 and 0.28 at t=0, 20, 60, 140, 200 orbits respectively. As these simulations are computationally expensive (on 8 cores it takes roughly a week to simulate 200 secondary orbits), we decided to use an open boundary condition where the eccentricity of the gas converges well within 200 orbits.

Mapping the parameter space in this way enables us to determine which parameter has a strong influence on the maximum particle size. 

\begin{table*}
\caption{Initial parameters and results of the FARGO and dust simulations.}             
\label{table:ini}      
\begin{center}                        
\begin{tabular}{c c c c c c c c c c c c c}        
\hline\hline                 
ID 	& $M_1$ 		& $M_2$ 		& e 	& a 		& $\alpha$	& $r_{in}$ 	& Inner b.&$t_{eq}$	& $\frac{<e \Sigma>}{<\Sigma>}$&$<m_{\mathrm{in}}>$	& $<m_{\mathrm{out}}>$& $<t_s>$ \\  
 	& [$M_{\odot}$]&[$M_{\odot}$] &  	& [AU]	& 			& [AU] 	&		&[orbits]	& &[g]	&[g] & \\    
\hline                        
`single' &1.4  		& 0  	 		& 0 	& 0		&$10^{-2}$ 	& 0.4	& non-refl.	&0		&0.004	&264		&446					&$7.9\times10^{-2}$\\ 
`no\_ecc' &1.4  		& 0.4   		& 0 	& 20		&$10^{-2}$ 	& 0.4	& non-refl.	&100		&0.012	&4.3		&$1.8\times 10^{-2}$		&$2.0\times10^{-2}$\\ 
`fiducial'  &1.4  		& 0.4   		& 0.4& 20		&$10^{-2}$ 	& 0.4	& non-refl.&100		&0.087	&1.6		&$4.8\times 10^{-4}$		&$4.6\times10^{-3}$\\ 
`openb'   &1.4  		& 0.4   		& 0.4& 20		&$10^{-2}$ 	& 0.4	& open 	&100		&0.045	&2.8		&$4.0\times 10^{-4}$		&$5.5\times10^{-3}$\\ 
`alpha'  &1.4  		& 0.4   		& 0.4& 20		&$10^{-3}$ 	& 0.4	& open 	&200		&0.126	&4.5		&$2.4\times 10^{-3}$		&$8.1\times10^{-3}$\\
`r\_in'   &1.4  		& 0.4   		& 0.4& 20		&$10^{-2}$ 	& 0.1	& non-refl.	&100		&0.087	&6.1		&$6.4\times 10^{-4}$		&$5.2\times10^{-3}$\\ 
`sep'	   &1.4  		& 0.4   		& 0.4& 15		&$10^{-2}$ 	& 0.4	& non-refl.	&200		&0.059	&0.88	&$7.7\times 10^{-7}$		&$2.4\times10^{-3}$\\ 
`sec\_m'&1.4  		& 0.6   		& 0.4& 20		&$10^{-2}$ 	& 0.4	& non-refl.	&100		&0.093	&0.48	&$3.8\times 10^{-4}$		&$1.8\times10^{-3}$\\ 
\hline
\end{tabular}
\end{center}
In this table, Col. 1 describes the identifier of the simulation pointing out the parameter which is different compared to the `fiducial' simulation (e.g., `alpha' means that the turbulence parameter ($\alpha$) is different than in the `fiducial' simulation). Cols. 2 and 3 are the masses of the primary and secondary respectively, Cols. 4 and 5 are the eccentricity and semi major axis of the secondary respectively, Column 6 is the turbulence parameter ($\alpha$), col. 7 is the radius of the inner boundary. Column 8 describes the boundary condition applied at the inner boundary, it can be non-reflective or open. Column 9 is the number of secondary orbits simulated before the equilibrium of the gas is reached, col. 10 is the surface density weighted average eccentricity of the gas at the end of the simulations. The properties of the final dust populations using the erosion model is described in cols. 11-13. Column 11 is the average mass of the particles between 0.5 and 1 AU, col. 12 is the average mass of particles between 4.5 and 5.5 AU, col. 13 is the average stopping time throughout the disk.

\end{table*}

\subsubsection{Initial conditions}
\label{sec:inicond}
We use 240 logarithmically spaced radial bins and 504 azimuthal bins to simulate the disk, resulting in nearly quadratic gridcells. It extends from $r_{in}$ = 0.4 AU until $r_{out}$ = 8 AU. In the case of the `r\_in' simulation, the radial bin goes from 0.1 AU until 8 AU and we use 351 logarithmically spaced radial bins to keep the size-ratio of the grids constant. The initial surface density profile is $\Sigma =\Sigma_0 r^{-0.5}$, the surface density at 1 AU is $\Sigma_0 = 250$ g/cm$^2$, which translates to an initial disk mass of $2\times10^{-3}$ M$_1$. The aspect ratio of the disk ($h$) is chosen to be 0.05. The temperature profile is $T(r) \propto r^{-1}$, which is the result of the assumed constant aspect ratio and it is kept unchanged during the simulations (no energy equation is solved).

Due to the mass loss during the initial stages of the simulation, the disk mass is roughly $1.5 \times 10^{-3}$ M$_1$, when the quasi-equilibrium is reached. Mass loss occurs in all simulations where a secondary is present, however it is more severe for an eccentric secondary. If the secondary is on a circular orbit, the eccentricity of the gas disk is increased in the outer disk (see Fig. \ref{fig:sigma0}b). The reason for mass loss is the following. Let us consider a fluid parcel that is initially in a circular orbit in one of the outermost grids of the simulation. Due to the perturbation of the secondary, this fluid parcel is forced to an elliptic orbit. As the apoastron of the fluid parcel is now greater than its original orbital distance, the fluid parcel leaves the computational domain through the open boundary and mass loss occurs. The problem is more pronounced for an eccentric secondary. Figure \ref{fig:sigma1}b shows the eccentricity of the gas disk when the secondary is in apoastron. However, the gas eccentricity in (and shortly after) periastron is above unity, therefore the gas in the outer disk is on a hyperbolic orbit and the gas disk between 6 and 8 AU is practically depleted (Fig. \ref{fig:sigma1}a).

\subsubsection{Boundary conditions}
We use most of the time the so called non-reflective boundary condition at $r_{in}$ which efficiently removes any reflected waves from the inner boundary \citep{Crida2008}. At every time step, the density in the zeroth ring is set to the density of the first ring, which is rotated to simulate wave propagation and avoid wave reflection. The open inner boundary condition lets matter freely flow out from the grid.

We always use an open boundary condition at $r_{out}$, which lets the matter leave the computational domain during the initial stages of the simulation. This boundary condition sets the radial velocity to zero, if it points inwards. The radial velocity is extrapolated otherwise and the matter can flow out. We also set a density floor in the outer ring, which we choose to be 10 orders of magnitude smaller than the average initial density. As explained in the previous Section, the eccentricity in the outer disk is above 1 when the secondary is in periastron. Therefore matter is continuously lost through the outer boundary and the density is decreasing after every passage of the secondary. We find that when the density is below a critical value (roughly 15-20 orders of magnitude below the average density), negative density values occur due to numerical errors. The density floor criteria ensures that this does not happen and its effect on the surface density profile is undetectable.

\subsection{Quasi-equilibrium of the gas disk}
The structure of a disk around a single star is different from a disk which is subject to the perturbation of a close secondary, therefore we first follow the relaxation of the disk into a new quasi-equilibrium before modeling the motion of the particles in a binary system. 

Figures \ref{fig:sigma0}a and b show the surface density profile and eccentricity profile of the gas at different times in the `no\_ecc' simulation respectively. We see that the disk material restructures, the outer radius of the disk becomes 7 AU after 100 orbits. We do not see strong waves (spiral arms) in the disk, we see however that the eccentricity of the disk increases towards the outer part of the disk.

The eccentric secondary excites two spiral waves whenever it passes periastron. The waves propagate all the way to the inner disk. During the initial orbits of the secondary, the spiral arms carry a large amount of matter outside the computational domain which is then lost from the system due to the increased gas eccentricity at periastron. Between two periastrons the disk becomes more circular, it relaxes. The disk radius is truncated to approximately 5-6 AU as shown in Fig. \ref{fig:sigma1}a for the `fiducial' simulation and the disk material restructures to a new equilibrium after 60 orbits. The azimuthally averaged eccentricity of the gas is presented in Fig. \ref{fig:sigma1}b. The average value is 0.09. The eccentricity increases between $r=4$ and 8 AU due to the stronger perturbations from the secondary. Table \ref{table:ini} shows the required number of orbits to simulate in order to reach equilibrium, the gas and the surface density weighted average eccentricity at the end of the simulations. The eccentricity of the gas is low, if the companion is on a circular orbit. The open inner boundary reduces the eccentricity by 50\%, the lower $\alpha$ value increases the eccentricity by 50 \%. The smaller inner boundary radius has no effect on the eccentricity. It is, however, interesting that a closer companion reduces the eccentricity of the gas compared to the `fiducial' simulation. 

We note that although we mostly use a non-reflecting boundary condition, we do not favor this boundary condition over any other possibility. The exact amount of mass deposited from the disk onto the star is determined by the interaction of ionized gas at several stellar radii from the star and the magnetic fields. It is far from clear what boundary condition should be used in a hydrodynamical simulation where the (simulated) inner disk extends to ~0.1 AU only. As an example \cite{Kley2008} uses four different types of boundary conditions and concluded that the different boundary conditions result in different disk properties. This is what we also observe.
The built in boundary conditions of FARGO are rigid, open and non-reflecting. The rigid boundary condition introduces reflected waves, the fully open boundary condition results probably in too high accretion rates. Therefore we prefer the non-reflecting boundary condition, but our choice is not based on more solid arguments than the above. A more thorough approach would be to explore the effects of all boundary conditions. However, we concentrate on the dust physics in this paper, we do not feel that such an approach is necessary. 

\subsection{Dust}
In this Section we review the equations of motion of a dust particle, the numerical method used to solve these equations and we present the tests we performed to verify the validity of our scheme. We also describe the erosion model used in our model. The full collision model, including growth, is described in Sec. \ref{sec:comp_coll_mod}.
\subsubsection{Equations of motion}
A dust particle feels the gravitational force of the primary and secondary stars, the aerodynamical drag and inertial forces. The particles considered in this work have negligible masses compared to the masses of the binary stars, thus we can use the equations of the restricted three-body problem. The equation of motion of a dust particle in a spherical coordinate system (r, $\varphi$) is
\begin{equation}
\frac{dr}{dt}=v_r
\label{eq:r}
\end{equation}
\begin{equation}
\frac{d\varphi}{dt}=L/r^2
\label{eq:phi}
\end{equation}
\begin{equation}
\frac{dv_r}{dt} = \frac{L^2}{r^3} - \frac{\partial \Phi}{\partial r}+F_{d,r}+F_{in,r},
\label{eq:vrad}
\end{equation}
\begin{equation}
\frac{dL}{dt} = - \frac{\partial \Phi}{\partial \varphi}+F_{d,{\varphi}}+F_{in,{\varphi}}, 
\label{eq:vthet}
\end{equation}
where $v_r$ is the radial velocity of the particle, $L$ is the specific angular momentum, $\Phi$ is the common gravitational potential of the two stars ($\Phi = -GM_1/r - GM_2/r_s$, where $r_s$ is the distance from the secondary star), $F_{d,r}$ and $F_{d,{\varphi}}$ are the radial and azimuthal drag forces respectively (see description in the next paragraph), and $F_{in,r}$ and $F_{in,{\varphi}}$ are the radial and azimuthal inertial forces, respectively (see description in a later paragraph). 

The drag force is calculated as
\begin{equation}
\bar{F_d} = - \frac{\Omega_k}{t_s}\Delta \bar{v},
\end{equation}
where $\Omega_k$ is the Kepler frequency, $\Delta \bar{v} = \bar{v}_d -\bar{v}_g$ is the relative velocity between the dust and the gas and $t_s$ is the stopping time of the particle. The stopping time (or friction time) is the time the particle needs to react to the changes in the motion of the surrounding gas. We follow the description used already by \cite{Paardekooper:2007p66}, \cite{Lyra2009a} or \cite{Woitke:2003p69} to calculate $t_s$. In this description, the stopping time for particles of intermediate sizes is calculated as an interpolation between the Epstein regime (for particles smaller than the mean free path of molecules in the gas \cite{Epstein1924}), and the Stokes regime (for particles much bigger than the mean free path \cite{Weidenschilling1977}). We refer the reader to the above mentioned papers for further details of the calculation. 

The inertial forces arise since the coordinate system in FARGO is centered in the primary star and not in the center of mass of the system. Therefore, the center of mass orbits around the primary star. The inertial forces are calculated in Cartesian coordinates for simplicity and later on they are transformed into spherical coordinates. The x and y components of the inertial forces in such a system are
\begin{equation}
F_{in,x}=\ddot{R_c} \cos{\Phi_c}+2\dot{R_c}\dot{\Phi}_c\sin{\Phi_c}-\dot{\Phi}_c^2R_c\cos{\Phi_c}- \ddot{\Phi}_cR_c\sin{\Phi_c},
\label{eq:inx}
\end{equation}
\begin{equation}
F_{in,y}=\ddot{R_c} \sin{\Phi_c}-2\dot{R_c}\dot{\Phi}_c\cos{\Phi_c}-\dot{\Phi}_c^2R_c\sin{\Phi_c}+ \ddot{\Phi}_cR_c\cos{\Phi_c},
\label{eq:iny}
\end{equation}
where $R_c$ and $\Phi_c$ are the (time dependent) distance and angle of the center of mass relative to the primary star. In case the secondary star is on a circular orbit (that is $R_c(t)=const.$ and $\Phi_c(t)=\omega t$), the inertial forces are reduced to $F_{in,x}=-\omega^2 R_c\cos{\Phi_c}$ and $F_{in,y}=-\omega^2 R_c\sin{\Phi_c}$, which have the same form as the centrifugal force expression. In the case of an elliptic secondary, the full expressions in Eqs. \ref{eq:inx} and \ref{eq:iny} must be used.

\begin{figure*}
  \includegraphics[width=0.5\textwidth]{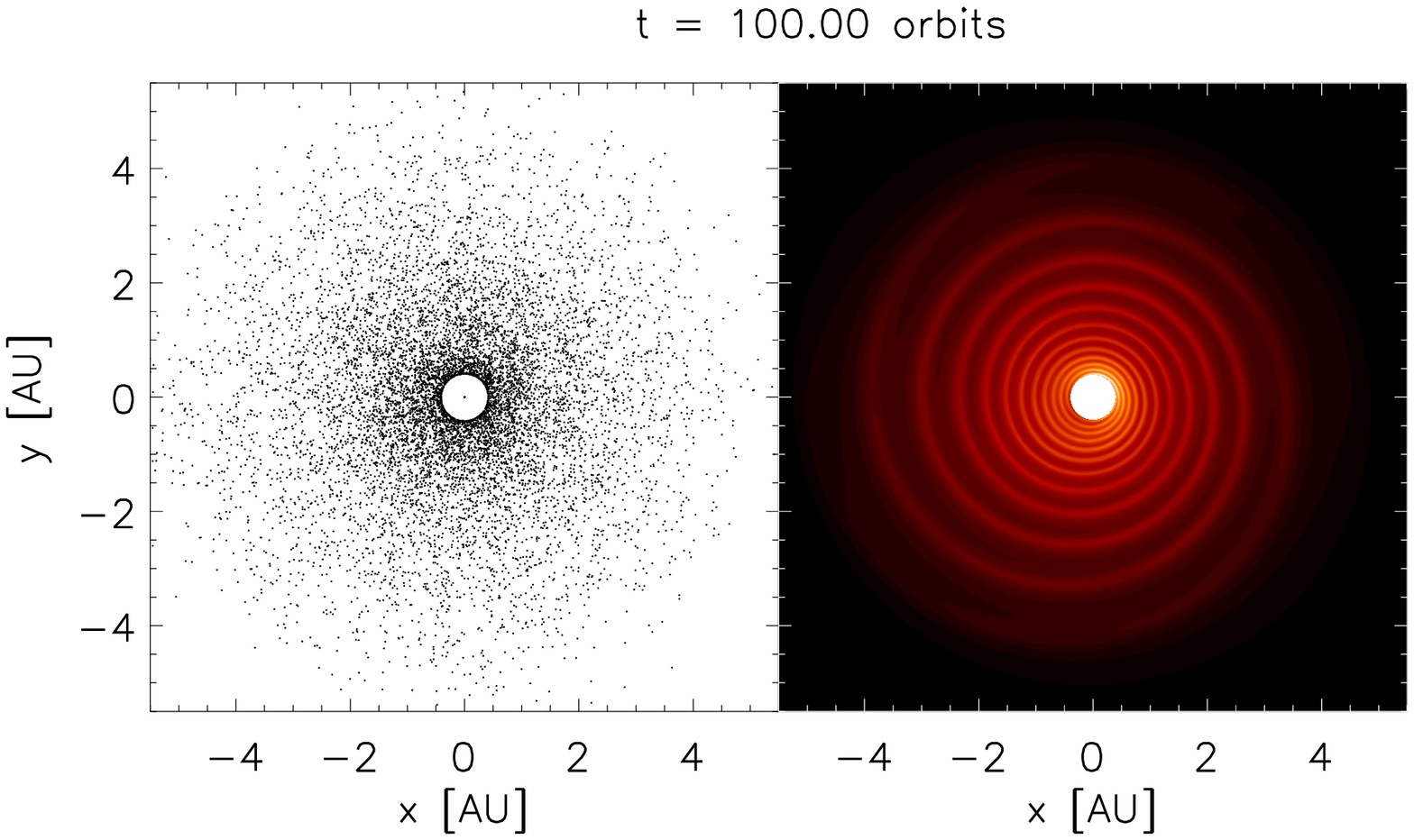}
  \includegraphics[width=0.5\textwidth]{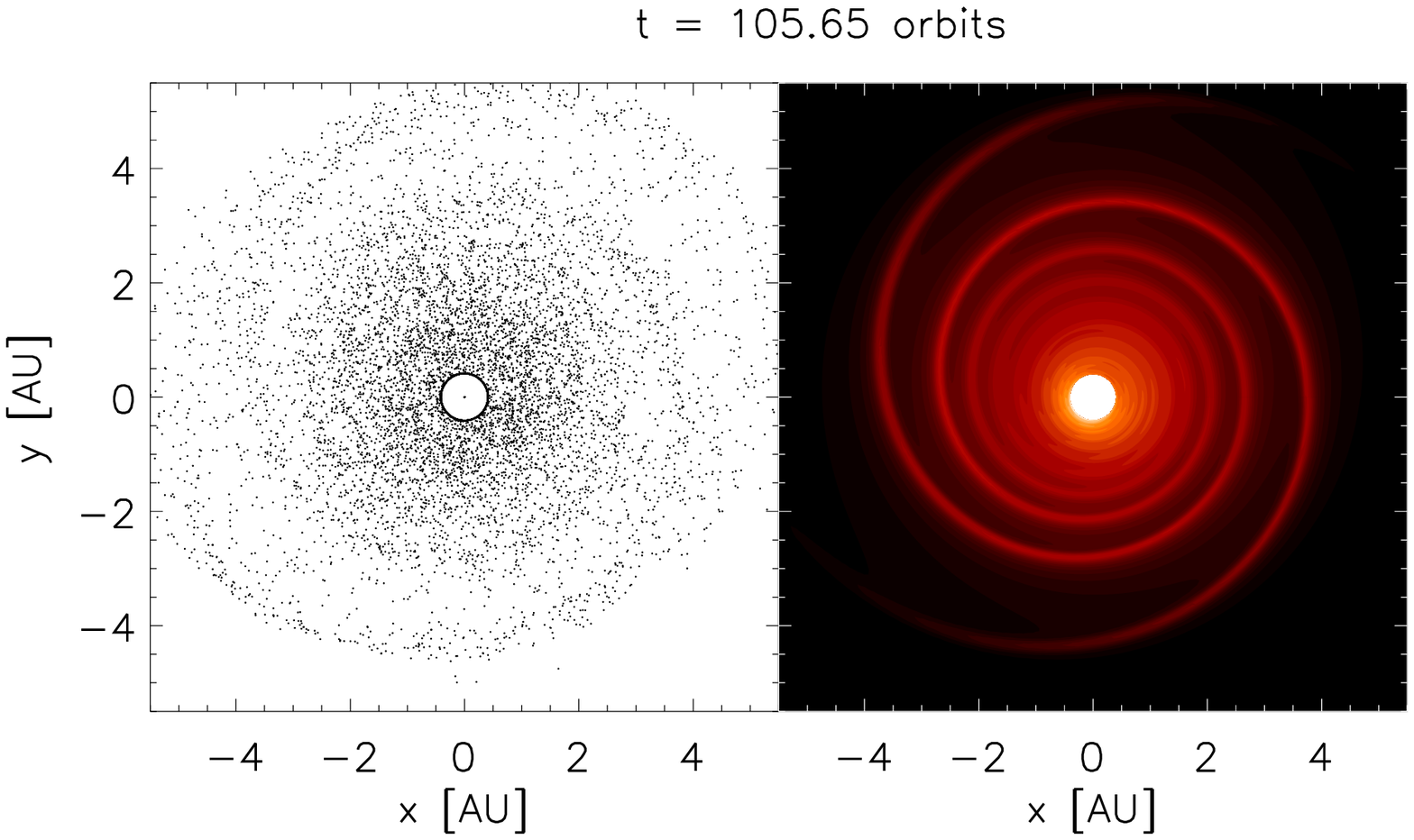}
  \caption{The surface density of the gas (contour figures on the right sides) and the dust (left sides) initially at $t=100$ orbits (a) and after $t=105.6$ orbits of the companion (b). The spiral waves in the gas as well as in the dust are clearly visible on Fig. b.}
  \label{fig:gtod_test}
\end{figure*}

\subsubsection{Numerical method}
We use a standard first order leapfrog integrator similar to that used by \cite{Paardekooper:2007p66}. The discretized forms of Eqs. \ref{eq:r}-\ref{eq:vthet} upon solving the drag force implicitly are
\begin{equation}
r=r_0+v_{r0}\Delta t
\end{equation}
\begin{equation}
\varphi = \varphi_0+L_0\Delta t / r_0^2
\end{equation}
\begin{equation}
v_r = \frac{v_{r0} + \Delta t \left( \frac{L^2}{r^3} - \frac{\partial \Phi}{\partial r} + \frac{v_{\mathrm{g,r}}\Omega_k}{t_s}+F_{in,r} \right)}{1+\frac{\Delta t \Omega_k}{t_s}},
\end{equation}
\begin{equation}
L = \frac{L_0 + \Delta t \left( - \frac{\partial \Phi}{\partial \varphi} + \frac{v_{\mathrm{g,\varphi}} r \Omega_k}{t_s}+F_{in,{\varphi}} \right)}{1+\frac{\Delta t \Omega_k}{t_s}},
\end{equation}
where $\Delta t$ is the time step, $v_{\mathrm{g,r}}$ and $v_{\mathrm{g,\varphi}}$ are the radial and azimuthal gas velocities respectively. Solving the drag force implicitly enables us to simulate the motion of an arbitrary small particle. We obtain the gas density and velocity values from the FARGO code. The gas frames are saved 100 times during one orbit of the companion, but the time step of our method is much smaller than that (usually $10^{-3}$ yr). Therefore we linearly interpolate between two neighboring time frames ($t_i$ and $t_{i+1}$) to obtain the gas properties at an intermediate time of the simulation $t_{\mathrm{sim}}$, where $t_i < t_{\mathrm{sim}}< t_{i+1}$. To determine the exact gas density and velocity components at the position of a particle, we perform bilinear interpolation using the four neighboring values. 

\subsubsection{Initial and boundary conditions}
The particles are distributed randomly in the disk in a way that the initial surface density profile of the dust is the same as the gas surface density profile. The particles have Kepler velocities initially. The boundary conditions for the dust is open both at the inner and at the outer boundary. If a particle leaves the FARGO grid, the particle is removed from the simulation. This differs from the gas boundary conditions which leads to some discrepancy between the surface density profile of the gas and small ($t_s \sim 0$) particles (see Sect. \ref{sec:tests}). 

\subsubsection{Tests}
\label{sec:tests}
We performed three tests to ensure that our numerical method is correct.

\paragraph{Stopping time of particles equals zero.} In this case the particles are well coupled to the gas. Thus the gas to dust ratio has to be constant during the simulation. We placed $10^5$ particles with $t_s = 0$ randomly in the hydro grid after the gas reached quasi-equilibrium (after $t=100$ orbits). The initial distribution of particles is the same as of the initial gas distribution (see Fig. \ref{fig:gtod_test}a). We follow the motion of these tracer particles for $334$ yrs (5 orbits) after which we compare the gas and dust distribution again (see Fig. \ref{fig:gtod_test}b). The tracer particles concentrate in the spiral arms similarly to the gas, but some differences can be observed. The inner part of the dust disk is depleted compared to the gas disk due to the different boundary conditions used for the gas and the dust. Although the inner radius of the gas disk is at 0.4 AU (unless otherwise stated), we only consider dust particles outside 0.5 AU. 

The different boundary conditions for the gas and the dust is justified because the dust behaves like a pressure-less fluid. Although distant gas parcels can communicate with each other via pressure effects, there is no interaction between dust particles at different parts of the disk (i.e. the motion of any dust particle is determined by its local gas density and velocity only). 

Other differences can be explained by the presence of shocks in the gas. Whenever shocks appear (secondary in periastron), jumps in the radial and azimuthal velocity is present. As we use bilinear interpolation to determine the gas velocity at the position of the dust particle, these jumps (shocks) are smeared out, therefore the trajectory of the tracer particles somewhat differs from the trajectory of the gas. Higher grid resolution helps this problem, but as the shocks cannot be resolved in hydrodynamical simulations, some difference between the gas and $t_s = 0$ particles is always present. As we are not interested in the evolution of the dust surface density or the gas to dust ratio, this effect is not a serious problem in our investigation. 

\begin{figure}
\centering
\includegraphics[width=0.5\textwidth]{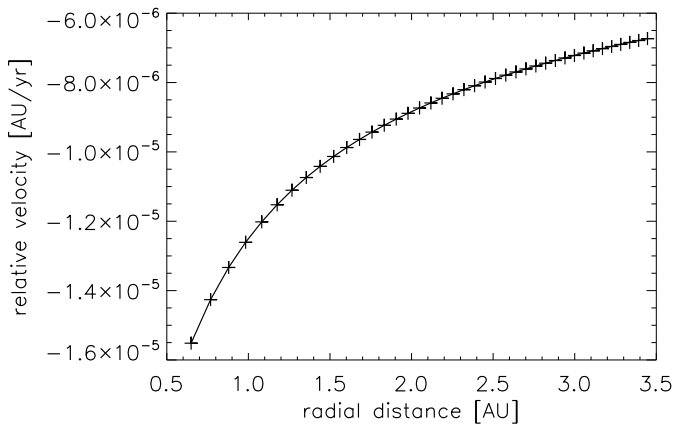}
\caption{Radial velocity difference between a dust particle with $t_s = 0.005$ and the gas. The solid line is the analytical drift speed, the `+' signs are obtained using our integration scheme.}
\label{fig:drift_test}
\end{figure}

\paragraph{Intermediate stopping time} We compare the drift speed of a test particle with the analytical results of \cite{Takeuchi:2002p73}. If the gas density profile is constant over the radius and the temperature profile follows a $T(r)\propto r^{-1}$, the radial drift velocity of the particle compared to the gas ($\Delta v_r$) can be given as
\begin{equation}
\Delta v_r = \frac{-h^2 v_k}{t_s+t_s^{-1}},
\label{eq:analdrift}
\end{equation}
where h is the aspect ratio of the disk, $h=0.05$ in our simulation. Fig. \ref{fig:drift_test} shows the drift velocity values for $t_s=0.005$ particles with the `+' signs, the analytical drift velocity dictated by Eq. \ref{eq:analdrift} is the solid line. The difference between the analytical and calculated drift speeds is at least six orders of magnitude smaller than the drift speed itself, therefore we conclude that our implicit integration scheme is correct and the obtained relative velocities are sufficiently accurate. 

\paragraph{Infinite stopping time} We placed a Jupiter mass planet initially at (1,0) coordinates on a circular orbit (no secondary star). In case of a particle with $t_s \gg  1$, the drag forces are negligible and the three body problem determines the motion of the particle. Thus a particle placed at the L4 or L5 Lagrange points should remain in the vicinity of these points and librate. Our test particle placed in the L4 point reproduces this movement.

\subsection{Erosion model}
\label{sec:erosion}
As described in the previous sections, we follow the motion of individual large dust aggregates, measure their velocity relative to the gas. Naturally, we can only integrate the motion of a limited number of particles. We assume that the rest of the dust population exist as ``background'' particles having small sizes ($\sim 1 \mu$m -- small dust population). This small dust population is very well coupled to the gas (their relative velocity is practically zero) therefore it is reasonable to assume that the relative velocity we measure is the collision speed between the large simulated particles and the small dust population. Our collision model is based on this assumption. 

One could also obtain the relative velocity between two larger aggregates, but one has to face some problems. The distance between two aggregates is not negligible. The Kepler shear, which is present on even sub-grid scales, introduces relative velocities which can easily be several tens of meters per second \citep{Lyra2009a}. One has to remove all such effects to obtain true collision speeds (e.g., relative velocity at zero distance). We discuss these effects in more detail in Sect. \ref{sec:critic}.
 
Our collision model includes erosion, if the measured relative velocity is greater than 1 m/s. First, we calculate how many micron-sized monomer (from the small dust population) collides with our aggregate during a time step. We calculate the aerodynamical cross section of the aggregate as
\begin{equation}
\sigma_{cr} = \pi r^2,
\end{equation}
where $r$ is the radius of the aggregate. We calculate the total gas mass that the aggregate swept through
\begin{equation}
m_g=V \times \rho_{\mathrm{gas}}=\sigma_{cr} \Delta v \Delta t \times \rho_{\mathrm{gas}},
\end{equation}
where $V$ is the gas volume the aggregate travelled through, $\Delta t$ is the time step of the calculation, and $\rho_{\mathrm{gas}}$ is the gas density. We assume a typical gas to dust ratio of 100 for the small dust population. Although we follow the motion of some individual large particles, their mass compared to the total dust mass is negligible, therefore we can assume that all dust mass is present in the form of micron sized monomers. The number of collisions during a time step is then given by
\begin{equation}
\label{eq:ncoll}
n_{\mathrm{coll}}=\frac{m_g/100}{m_0},
\end{equation}
where $m_0$ is the mass of a monomer. Based on the laboratory experiments with erosion of dust aggregates, the mass of the particle after a collision is \citep{Guttler2010, Wurm:2005p81}
\begin{equation}
\label{eq:truemassloss}
m_{\mathrm{new}}=m_{\mathrm{old}}-\frac{8}{60}m_0\frac{\Delta v}{100 \: \mathrm{cm/s}}.
\end{equation}
Using this equation, however, resulted in very slow mass loss, thus the integration of hundreds of secondary orbits would have been necessary to reach the final particle sizes. We therefore speed up erosion artificially, as we are only interested in the final particle sizes and not in the exact erosion timescales. Our erosion equation is
\begin{equation}
\label{eq:massloss}
m_{\mathrm{new}}=m_{\mathrm{old}}-10 m_0 \frac{\Delta v}{100 \: \mathrm{cm/s}}.
\end{equation}
We obtain final particle masses in less than 20 orbits using the equation above.

This is clearly a simplified collision model as in a realistic dust distribution, the collision partners can have higher masses, thus the number of collisions per time step is reduced. Also, the collision physics between similar sized particles is different (more about this in Sec. \ref{sec:critic}).

We note that after an erosion event, the cross section and the radius of the particle, and therefore the stopping time, the relative velocity and the number of collisions changes. If multiple erosion events happen during a single time step, these changes can be unresolved. To avoid such a numerical artifact, we use an adaptive time stepping scheme which reduces the time step by a factor of two, when the mass-decrease of the particle ($m_{\mathrm{new}}/\Delta m$) is greater than $10^{-3}$. Also, the time step is increased by a factor of two up to a maximum value of $10^{-3}$, if $m_{\mathrm{new}}/\Delta m$ is greater than $10^{-3}$. We find, however, that the time-step is never decreased during the simulations even when Eq. \ref{eq:massloss} is used.

In principle, a simulation would be completed when the mass of the dust particles do not change anymore. This condition could be fulfilled if the particles had a fixed orbit. However, due to radial drift, this is not the case. By performing numerical experiments, we find that after 20 orbits of the secondary, a mass distribution that only slowly varies due to radial drift is achieved, if the starting size (or mass) of the particles is not too far from the final dust mass. The initial dust mass therefore is not the same for all the simulations. In the `single', `no\_ecc' and `fiducial' simulations, the initial particle mass is 500 g, 100 g and 50 g respectively. As seen on Figs. \ref{fig:mass_dist0}, \ref{fig:mass_dist1} and \ref{fig:mass_dist2}, all particles lose mass by the end of the simulation, therefore the initial mass values are appropriate.

\section{Results}
\subsection{Dust in disks with and without a secondary}
\label{sec:res1}
We proceed in three steps in order to understand the effects of an eccentric binary on the dust population. First, we determine the properties of the dust particles in a disk around a single star (`single' set-up). Then, we compare these results with a disk that is restructured by a non-eccentric secondary (`no\_ecc' set-up), and finally, we examine the effects of an eccentric binary (`fiducial' set-up).

\begin{figure}
\centering
\includegraphics[width=0.5\textwidth]{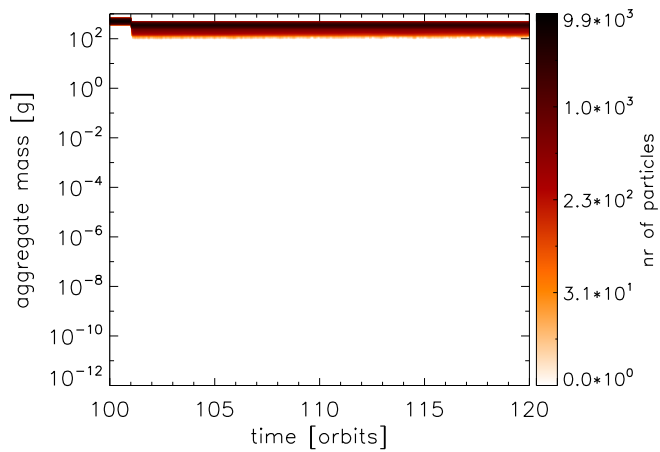}
\caption{The evolution of the mass of the particles in case of the `single' set up (no secondary present, the disk is not perturbed). The x axis is time in orbital periods of the secondary in the `no\_ecc' or `fiducial' set ups (one orbit takes 66.7 yrs). The y axis is the mass of the particles. The colors represent how many particles are present at a given location of the parameter space.}
\label{fig:mass_dist0}
\end{figure}

\begin{figure}
\centering
\includegraphics[width=0.5\textwidth]{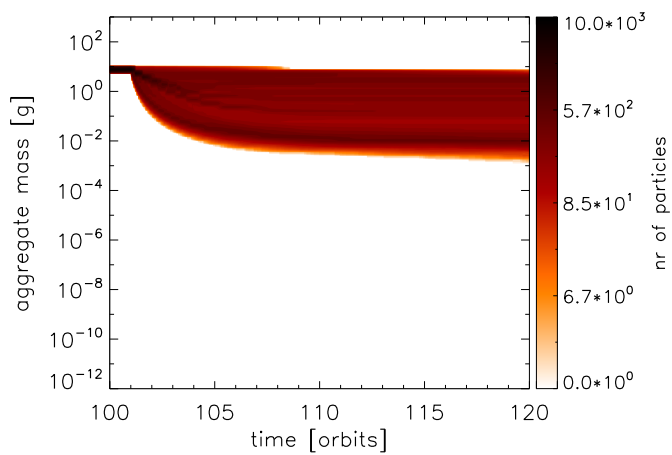}
\caption{Same as Fig. \ref{fig:mass_dist0} for the `no\_ecc' set up (a secondary is present on a circular orbit). The disk is truncated, which results in lower particle masses compared to the `single' simulation.}
\label{fig:mass_dist1}
\end{figure}

\begin{figure}
\centering
\includegraphics[width=0.5\textwidth]{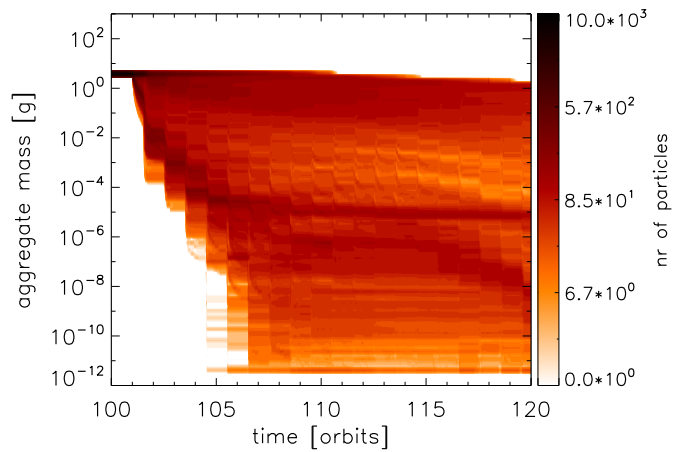}
\caption{Same as Fig. \ref{fig:mass_dist0} for the `fiducial' simulation, where a secondary is present on an eccentric orbit with $e=0.4$. The secondary closely approaches the disk at periastron (at every half orbital time) at which times the particles at the outer disk experience increased relative velocities, therefore they erode.}
\label{fig:mass_dist2}
\end{figure}

\begin{figure*}
  \includegraphics[width=0.5\textwidth]{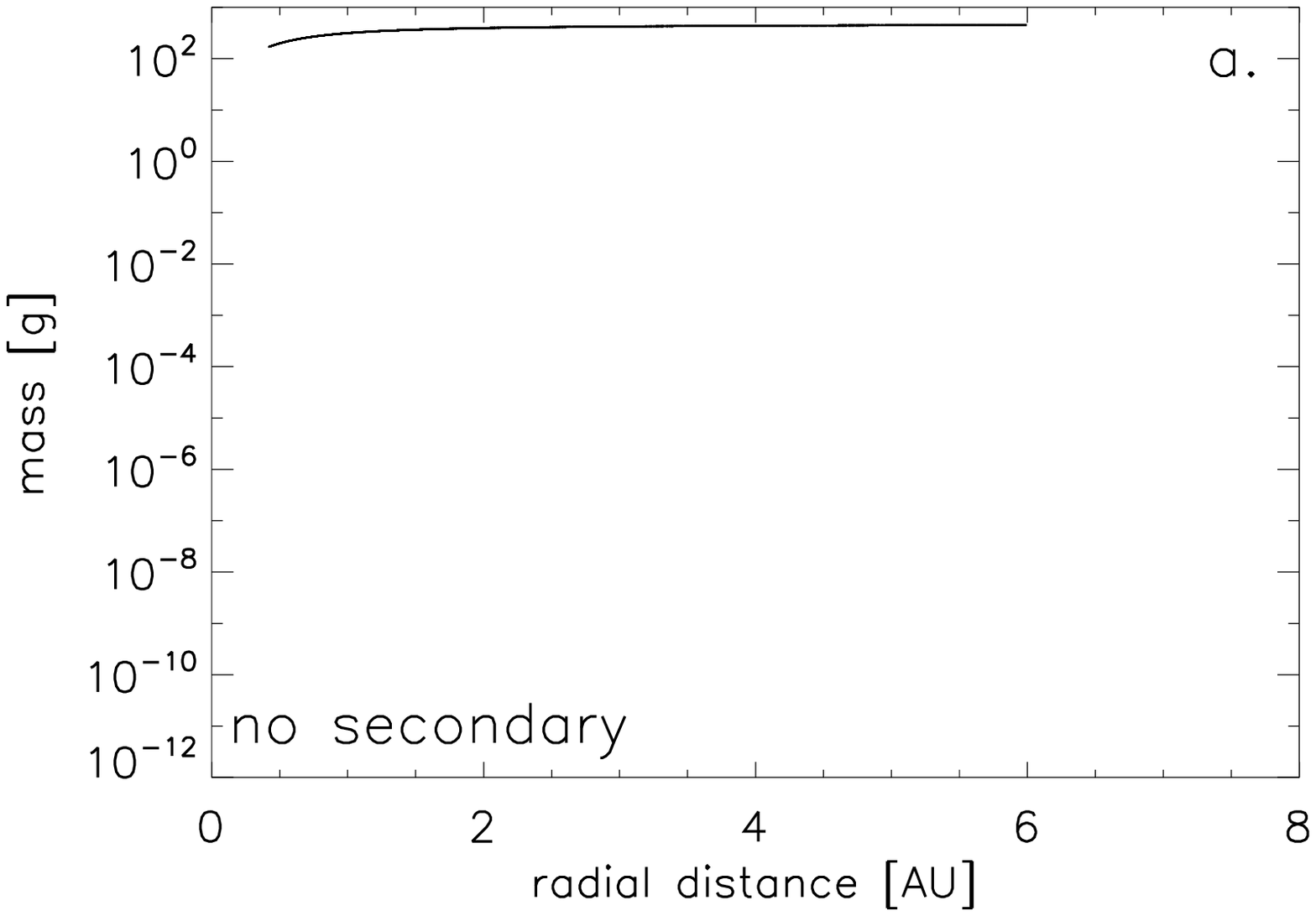}
  \includegraphics[width=0.5\textwidth]{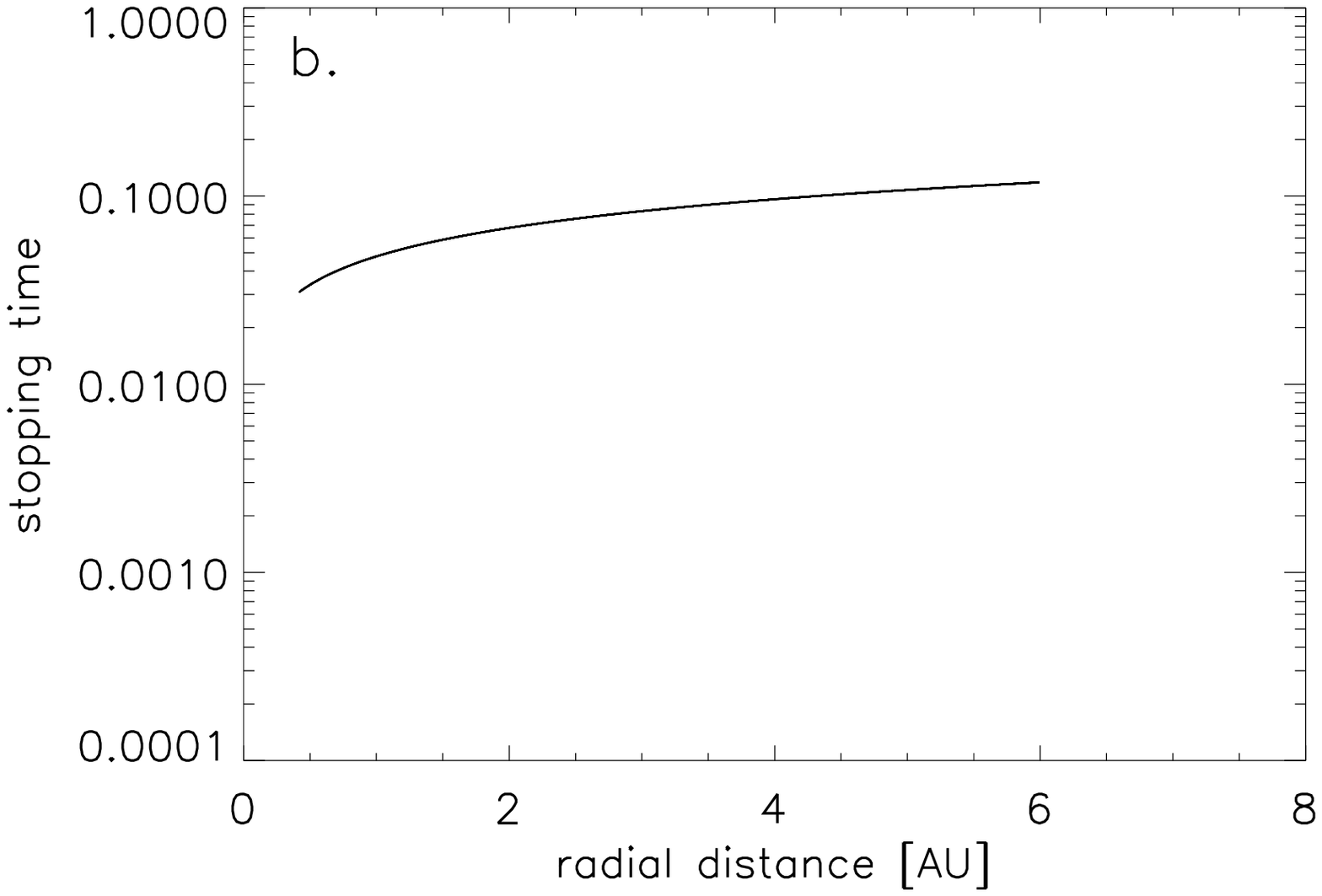}
  \includegraphics[width=0.5\textwidth]{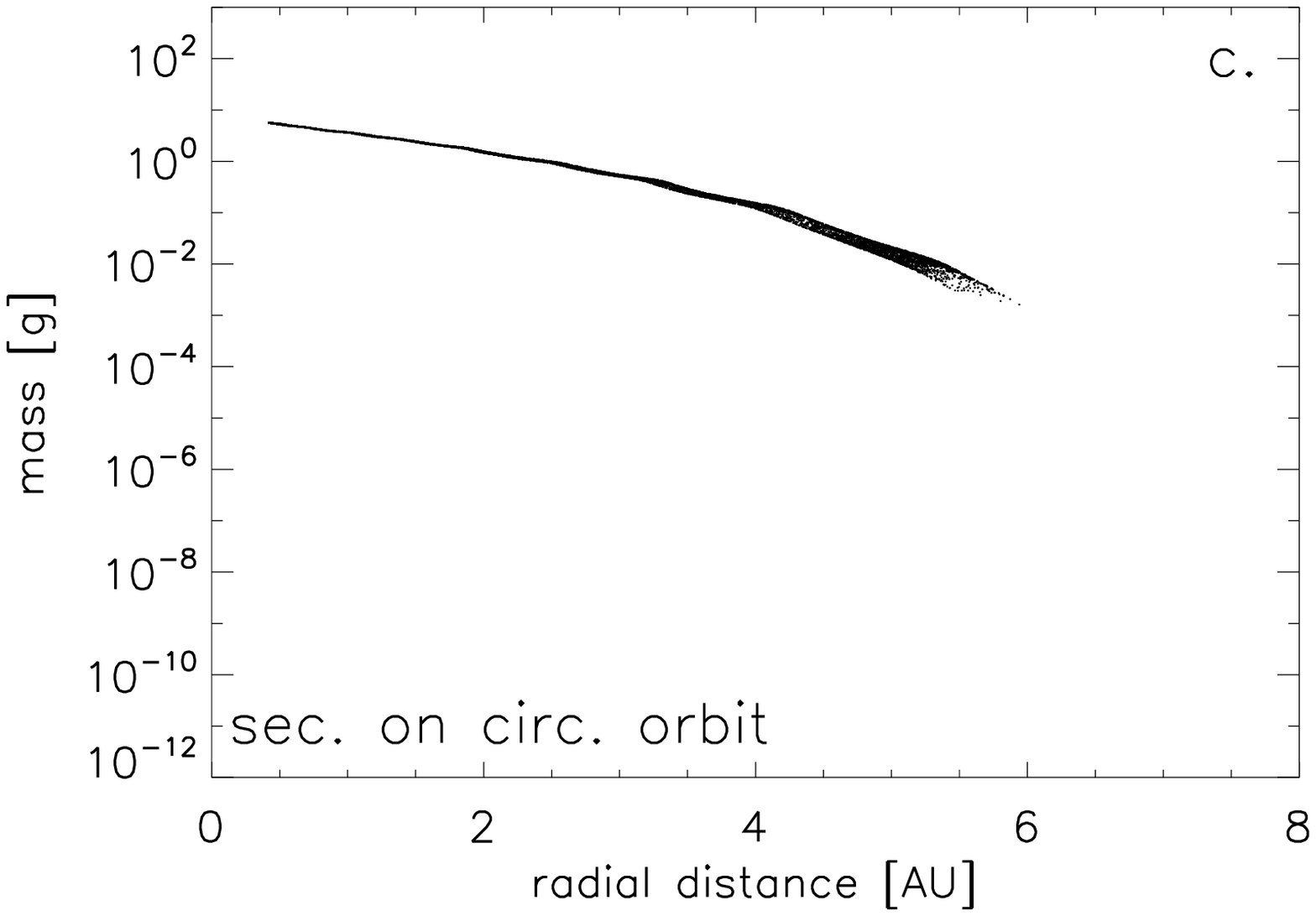}
  \includegraphics[width=0.5\textwidth]{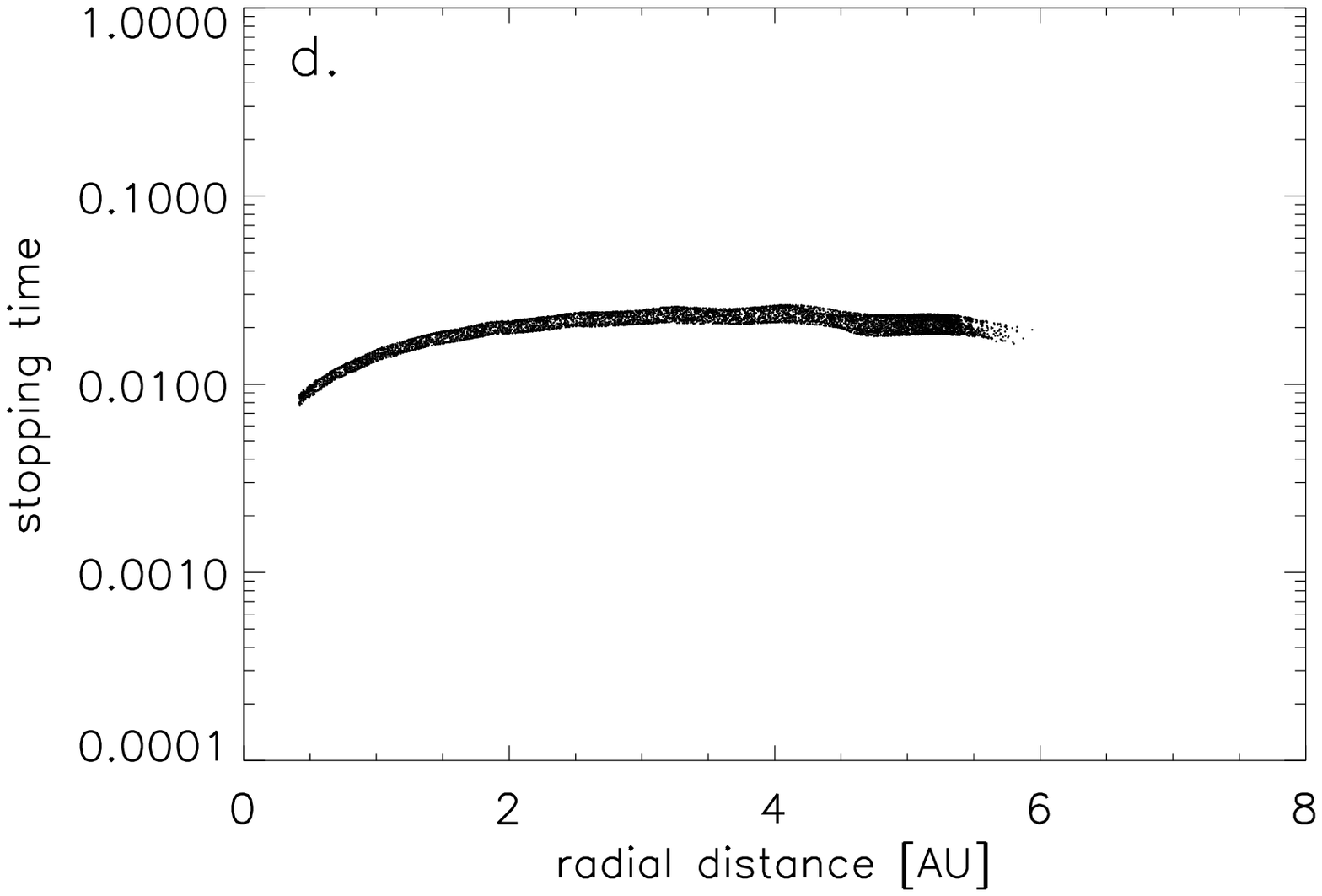}  
  \includegraphics[width=0.5\textwidth]{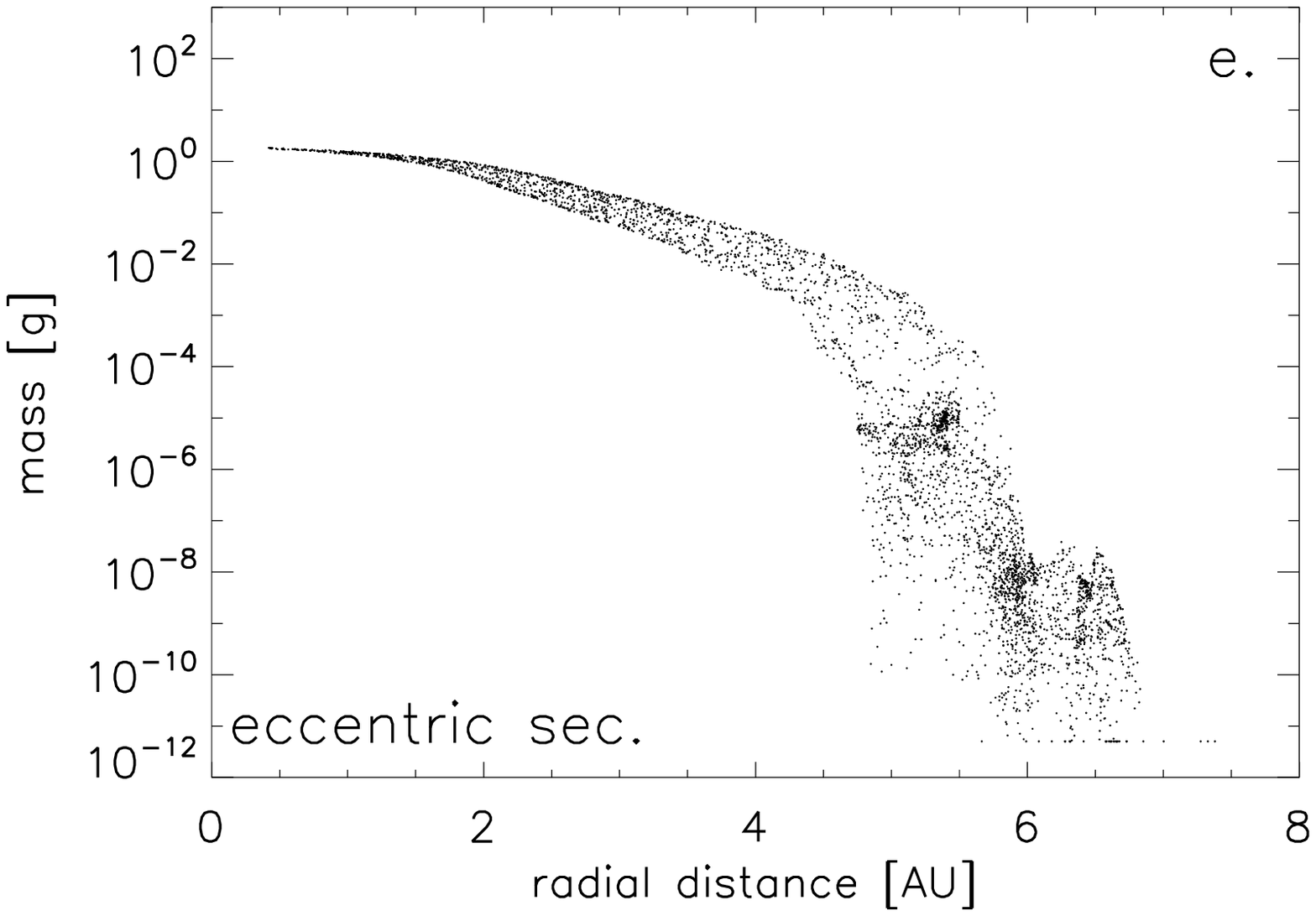}
  \includegraphics[width=0.5\textwidth]{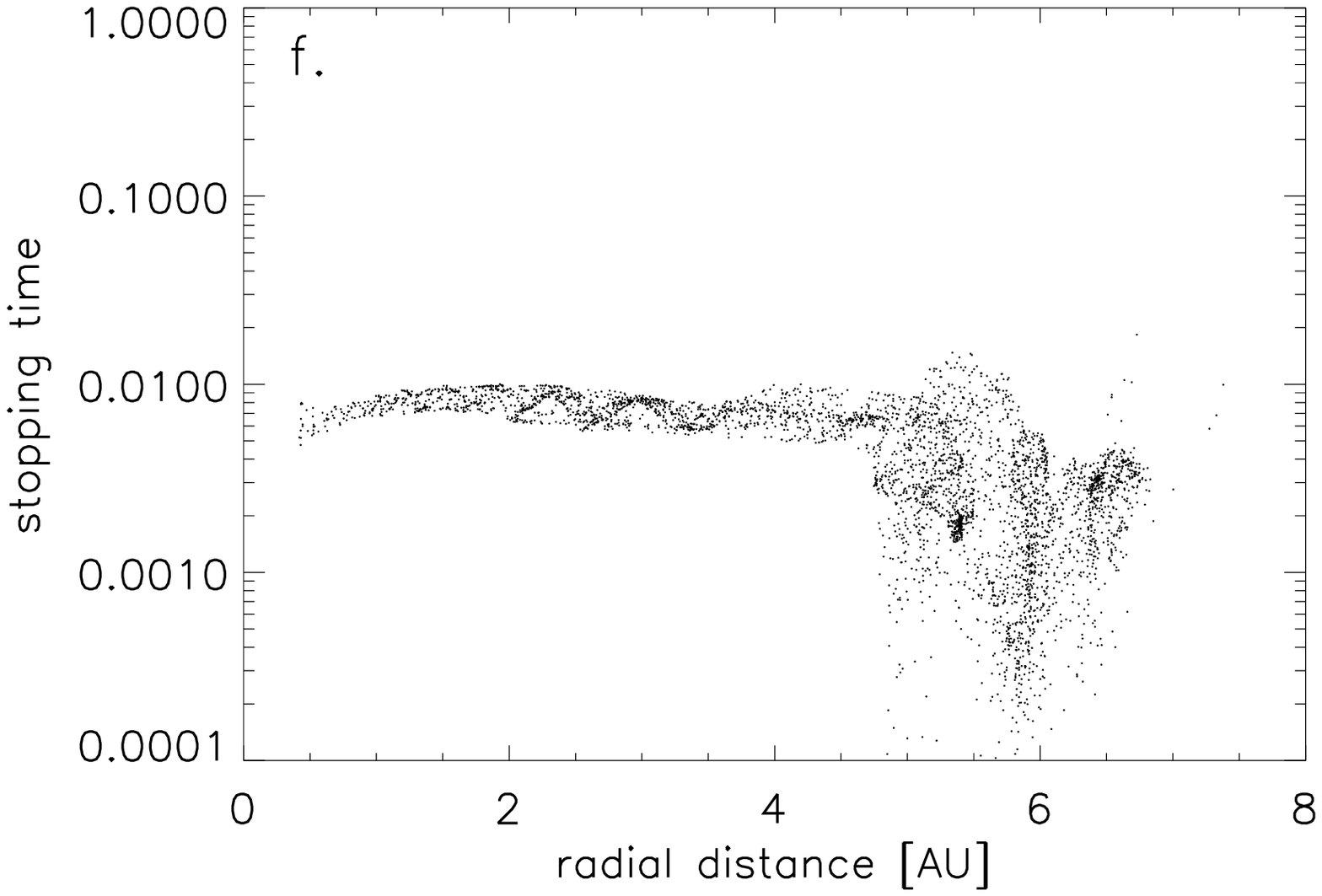}

  \caption{The figures on the left side (a, c, e) show the mass of the particles as a function of their distance from the primary at $t=120$ orbits for the `single', `no\_ecc', and `fiducial' simulations respectively. The figures on the right side (b, d, f) show the corresponding stopping times of the particles at the same time.}
  \label{fig:res1}
\end{figure*}

The gas density and velocity profile in the `single' model is the initial disk profile used for all of the simulations. That is a power-law disk profile, which is in hydrostatic equilibrium and has a constant accretion-rate throughout the disk (for details see Sec. \ref{sec:inicond}). We keep this density and velocity profile constant during the simulation of the particles. As shown in Tab. \ref{table:ini}, the eccentricity of the gas is the lowest of all set-ups (0.004). The small eccentricity is the result of the radial drift motion of the gas and it is also constant in time. The only source of relative velocity in this model is the differential radial drift of the dust. The dust simulation is started with particles having 500 g. We measure time in units of the orbital period of the secondary in the `no\_ecc' or `fiducial' models to be able to easily compare the results of the different models. The mass of the particles is kept constant for one orbital period (66.7 yrs) during which time the particles forget their initial Kepler velocity and sufficiently couple to the gas. After 66.7 yrs we turn on erosion and follow the mass-change of the particles as shown in Fig \ref{fig:mass_dist0} . Figures \ref{fig:res1}a and b show the final mass and stopping time of the particles as a function of their distance from the primary star respectively. The stopping time in general does not change strongly with the distance, therefore it can be characterized with the average. However, the mass of the particles can vary greatly with distance, thus we characterize the mass with its average around the inner and the outer edge of the disk. The average mass between 0.5 and 1 AU as well as between 4.5 and 5.5 AU, and average stopping time of the particles are shown in Tab. \ref{table:ini}.

A similar procedure is followed for the `no\_ecc' simulation, where a secondary is on a circular orbit. The disk is truncated, the outer edge of the disk is around 7 AU. The relative velocity in this, and all the other simulations is the result of both the radial drift and the ever-changing gas velocity profile. The particles have initially 10 g mass. The mass-evolution of the particles in this simulation is shown in Fig. \ref{fig:mass_dist1}. The final masses and stopping times of the particles as the function of their distance are shown in Figs. \ref{fig:res1}c and d respectively. The average mass at the inner, and outer disk as well as the average stopping time can be seen in Tab. \ref{table:ini}.

The mean free path of the gas atoms and molecules is of the order of 1 meter in these disk models due to the moderate densities. Furthermore the radius of the particles is below 10 centimeters. Therefore, we conclude that the particles are in the Epstein regime. The stopping time in this regime is 
\begin{equation}
t_{s} = t_{\mathrm{Ep}} = \frac{3 m}{4 v_{\mathrm{th}} \rho_g \sigma_{cr}},
\label{eq:epstein}
\end{equation}
where $v_{\mathrm{th}}$ is the thermal velocity of the gas. The relative velocity of the particle depends on how fast the particle can adapt to the changes in the gas velocity field, e.g., the relative velocity is higher, if the stopping time is higher. However, the gas density is decreased in a truncated disk. As seen in Eq. \ref{eq:epstein}, lower gas density increases the stopping time, thus the relative velocity is higher as well. Erosion, therefore, reduces the masses of the particles in the outer disk.

Therefore, one effect of the secondary is the reduced particle masses in the outer disk. The stopping time of the particles is also lower than in the `single' model. As seen in Tab. \ref{table:ini}, the eccentricity of the gas is three times higher in this simulation, therefore the particles, which would move on a circular orbit, experience stronger changes in the gas velocity field. Thus the stopping time is also reduced until the relative velocity reaches the critical 1 m/s value. The destructing effect of the gas eccentricity on the dust population affects the entire disk.

The particles in the `fiducial' simulation have initially 5 g masses. The mass evolution is the most dramatic in this case. The companion approaches the primary as close as 12 AU in periastron, therefore the outer edge of the disk experiences strong disturbances periodically. Every time this happens, the mass of the particles in the outer disk is reduced, which can be seen in Fig. \ref{fig:mass_dist2}. The secondary in periastron excites spiral arms in the gas disk as it is pushed inward. Therefore, a companion in an eccentric orbit further increases the eccentricity of the gas, thus decreasing the mass and the stopping time of the particles (see Tab. \ref{table:ini}, Figs. \ref{fig:res1}e and f). This effect influences the dust distribution in the whole disk. 

We conclude that the average stopping time is reduced by one order of magnitude in the `fiducial' simulation compared to the `single' simulation. The average mass of the particles however is roughly four orders of magnitude lower. 

\subsection{Exploring the parameter space}
We perform altogether eight simulations with different initial conditions as shown in Tab. \ref{table:ini}. We see that if the turbulent parameter is kept constant, different parameters (like the inner boundary condition, separation of the binary system, mass of the secondary) affect the gas, especially the eccentricity of the gas as shown in Tab. \ref{table:ini}. We perform similar dust simulations as the ones discussed in the previous section in all of these models and measure the average dust particle mass at the inner and outer disk, as well as the average stopping time of the particles at the end of the simulation. The results are shown in Tab. \ref{table:ini}. As a general trend, we can say that given a turbulence parameter, the average mass and stopping time of the particles decrease as the eccentricity of the gas increases. 

The `alpha' simulation with ten times lower $\alpha$ value produces the most eccentric disk, but the mass and stopping time of the particles are higher than the average of all the $\alpha=10^{-2}$ cases. The reason for this is that if the turbulence is lower, the radial flow of the gas is lower, the fluid behaves in a less viscous way. Therefore the dust relative velocities are also decreased allowing heavier particles to exist in such a disk.

The radius of the inner disk in the `r\_in' simulation is close to the usually assumed sublimation limit of the dust (where the temperature is higher than 1500 K). We performed this special simulation to see whether the effects of the secondary reaches the very inner parts of the disk. Previous work done on the growth of planetesimals in binary systems showed that the planetesimals in the inner regions are less affected by the secondary star, therefore these regions are favorable for their growth ( see e.g., \cite{Th'ebault2004}). In contrast, the growth of dust aggregates is less responsive to the direct gravitational perturbation of the secondary, but reacts sensitively to the gas dynamics. As the eccentricity of the gas is pumped up by the secondary throughout the disk, even the close regions to the primary star do not favor dust growth.

\begin{figure}
\centering
\includegraphics[width=0.5\textwidth]{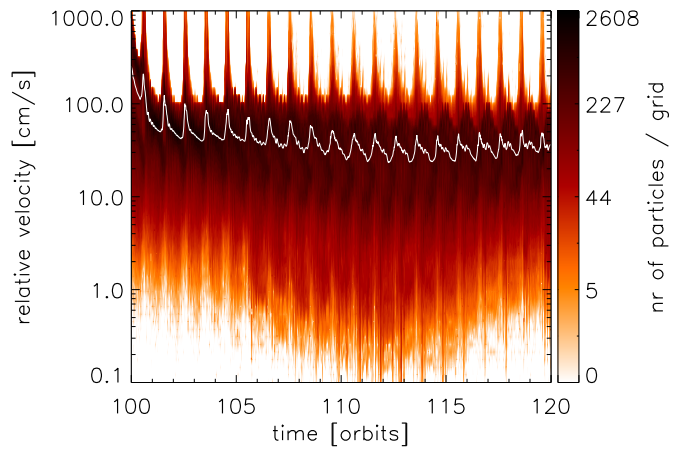}
\caption{The relative velocity distribution of aggregates as a function of time in the `fiducial' simulation. The x axis is in orbital time units, the y axis is the relative velocity in cm/s. The colors indicate how many particles are present at a given location of the parameter space. The white solid line shows the average relative velocity of all particles as a function of time.}
\label{fig:vrel_dist}
\end{figure}

\subsection{Long term prediction}
\label{sec:longtp}
The erosion model alone can tell us the maximum particle sizes in the disk that does not experience any fragmentation anymore. In the case of the `fiducial' set-up, however, the relative velocity of the particles is only increased when the secondary is in periastron as shown in Fig. \ref{fig:vrel_dist}. The relative velocity between two periastrons is much smaller than the critical velocity. The particle could experience growth during these periods, which alters the equilibrium mass distribution of the particles. 

To examine the long term evolution of the dust and calculate the equilibrium mass distribution as the function of the distance from the primary, we add coagulation to the collision model, distribute the particle masses randomly and follow the mass evolution of the particles for one orbit of the secondary. For more than one orbit, the radial drift of the large particles can be significant, thus non-local effects (due to radial drift) could alter the equilibrium mass distribution. Simulating less than one orbit is not sufficient since the appearance and dissipation of the spiral arms happens in an orbital period. We then calculate the growth/erosion rate of the particles ($(m_2-m_1)/m_1/T$, where $m_1$ is the starting mass, $m_2$ is the final mass of the particle in one orbit - $T$) as a function of their distance and stopping time and identify the zero erosion/growth rate curve. If we assume that the gas density and velocity profiles do not change significantly over multiple orbits, this curve identifies the equilibrium mass distribution between growth and erosion. In the following section, we describe the changes in the collision model, and the new initial conditions.

\subsubsection{The complete collision model}
\label{sec:comp_coll_mod}
In this section we introduce a new collision model that incorporates growth and erosion. Although we call this collision model complete, it does not incorporate all known collision types (see Sec. \ref{sec:critic}).

\paragraph{Growth} Equation \ref{eq:ncoll} describes the number of collisions which takes place during one time step of the calculation. According to the experiment-based collision model of \cite{Guttler2010}, it is realistic to assume that a monomer sticks to an aggregate of arbitrary mass, if the relative velocity is lower than the critical fragmentation velocity. Therefore, in the new collision model, $n_{\mathrm{coll}}$ number of monomers are added to the aggregate, if $\Delta v$ is less than 1 m/s. 

\paragraph{Erosion} In the previous collision model (see Sec. \ref{sec:erosion}), we artificially increased the erosion efficiency (see Eq. \ref{eq:massloss}) in order to obtain the final masses of the aggregates in a reasonable computational time. This is not necessary anymore, as we follow the mass evolution of particles only for one secondary orbit (see Sect. \ref{sec:longtp}). Therefore we use Eq. \ref{eq:truemassloss} to follow the erosion of the particles, if the collision velocity is greater than 1 m/s.

\subsubsection{New initial conditions}
In the following calculations the particles are initially uniformly distributed in radius between $r_{in}$ and $r_{out}$ (previously, the particles were distributed according to the gas surface density profile). Also, their masses are initially distributed uniformly in log-space between a monomer mass and 500 g. The initial velocity of the grains is Keplerian, and (as the stopping time of all particles is lower than the orbital time) for one obit we follow only the motion of particles to let the particles couple sufficiently to the gas. We turn on the collision model after one orbit of the secondary.

\begin{figure*}
  \includegraphics[width=0.95\textwidth]{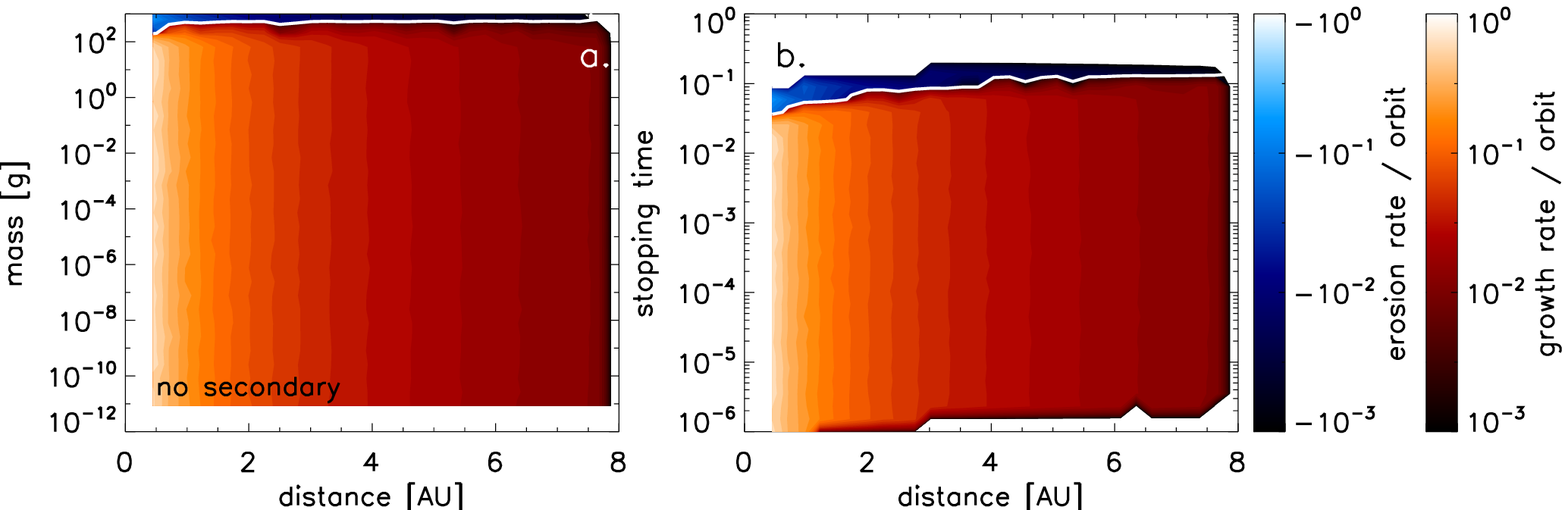}
  \includegraphics[width=0.95\textwidth]{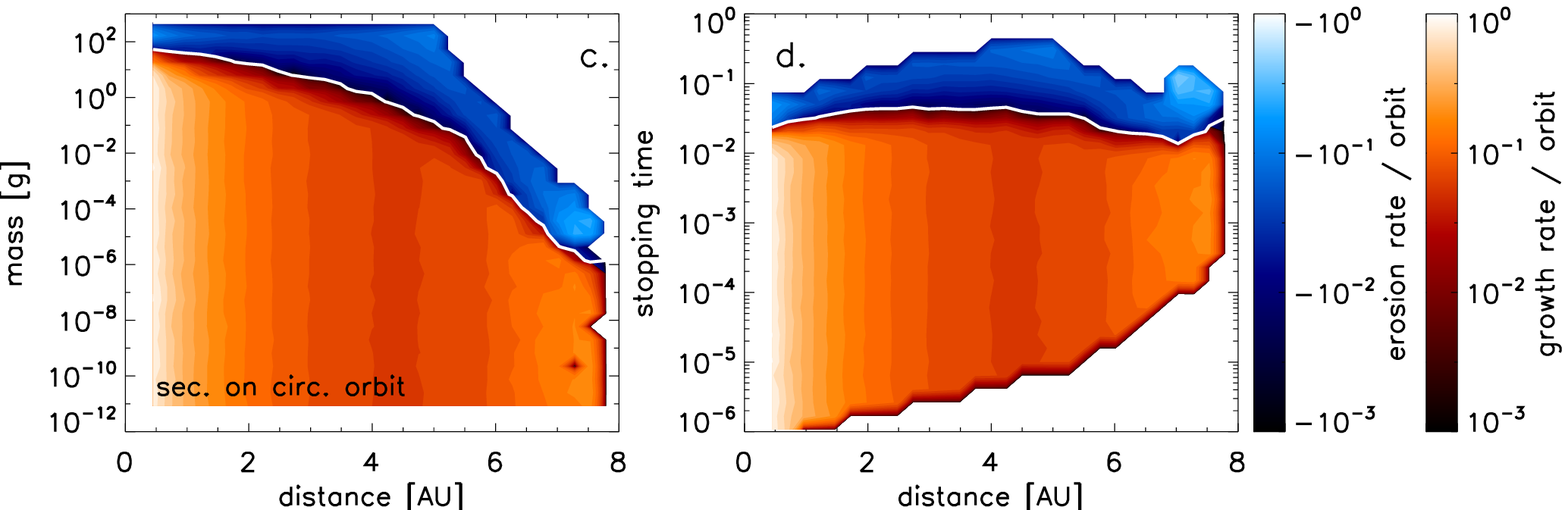}  
  \includegraphics[width=0.95\textwidth]{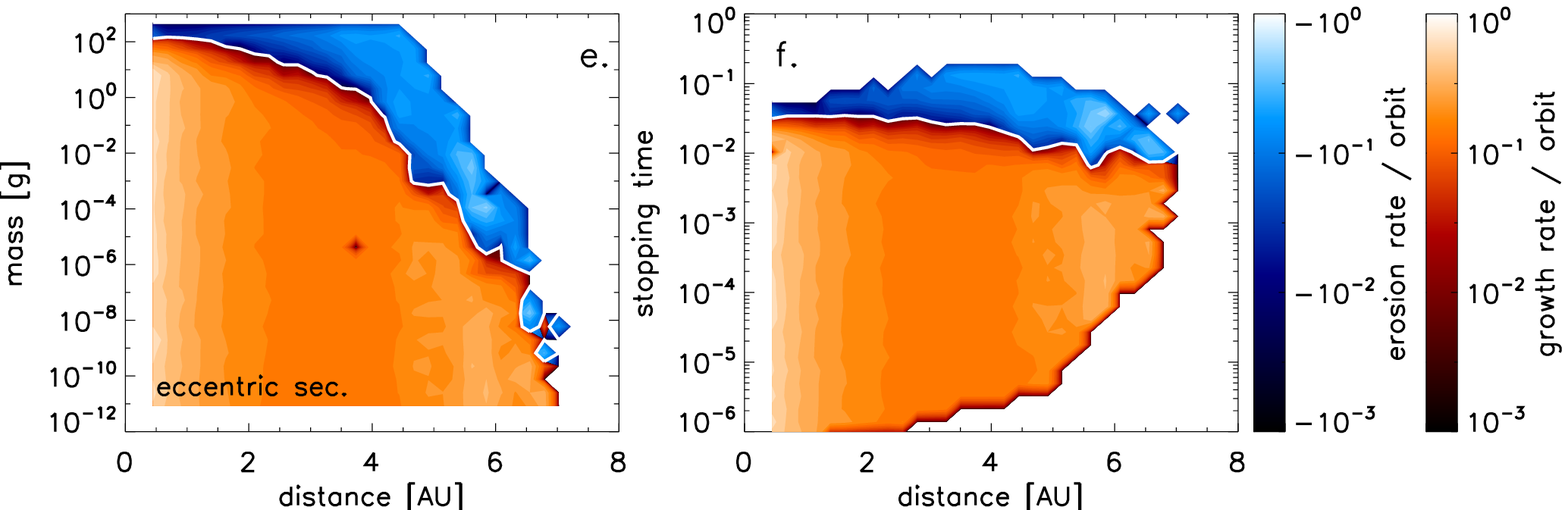}

  \caption{The figures on the left side (a, c, e) show the growth/erosion rates of particles in the `single', `no\_ecc' and `fiducial' simulations respectively. The x axis is the distance from the primary in AU units, the y axis is the mass of the particles in gram. The red contours (below the white line) illustrate the growth rate, white meaning that the particle doubled in mass during one orbit. The blue contours (above the white line) represents erosion rates, white in this case meaning that the particle completely eroded in one orbit. The figures on the right side (b, d, f) show the same quantities, but the stopping time is plotted on the y axis. The white solid line indicates the equilibrium between growth and erosion.}
  \label{fig:res2}
\end{figure*}

\subsubsection{Results}
Figures \ref{fig:res2}a and b, \ref{fig:res2}c and d, and \ref{fig:res2}e and f show the growth/erosion rates for the `single', `no\_ecc', and `fiducial' simulations, respectively. The growth/erosion rates are expressed in normalized dimensionless units ($(m_2-m1)/m_1/T$, where $m_1$ is the initial mass of the particle, $m_2$ is the final mass, and $T$ is the orbital period of the secondary). If e.g., the growth rate is unity, it means that the large aggregate doubled its mass during one orbital time of the secondary. 

Figure \ref{fig:res2}a and b for the `single' set-up shows that the growth/erosion rate decreases/increases with radius which is due to the decreasing density, thus lower number of collisions, as a function of distance.  

Figures \ref{fig:res2}c and d, and \ref{fig:res2}e and f show more complex features. The growth/erosion rates first decrease/increase as a function of distance, but it increases/decreases again at the outer parts of the disks. This behavior can be explained by the eccentricity of the gas. As seen in Figs. \ref{fig:sigma0}b and \ref{fig:sigma1}b, the eccentricity of the gas is higher in the outer parts of the disks due to the perturbation of the secondary. Due to the the eccentric gas at the outer disk, the relative velocities rise, therefore as long as the relative velocity is not above 1 m/s, the growth rate is increased again. 

We compare the equilibrium curve on these figures with the obtained solutions in Figs. \ref{fig:res1}a-f. The agreement is good for the `single' simulations. However, differences can be observed in the case of the `no\_ecc', and `fiducial' simulation and in all other simulation set-ups where the secondary is eccentric. The stopping time and mass in the updated collision model is generally higher in the outer disk than in the previous collision model including only erosion. The spiral arms excited by the secondary erodes the particles in periastron, but the aggregates can grow significantly between two periastrons. Therefore, calculating only erosion can underestimate the equilibrium size/stopping time of the particles.

Other difference can be observed in the inner disks. The mass and stopping time of the particles are higher in the full collision model than in the previous (erosion only) model. There are two reasons for that. First, the particles in the previous model do not grow between two periastrons. Secondly, the radial drift of the particles become important so close to the inner disk edge. As we only follow the particle motion for one orbit in the full collision model, the drift of the particles is negligible. However, in the previous model, the motion and dust evolution of particles is followed for 20 orbits (roughly 1500 yrs), therefore many particles in the inner disk with masses higher than 10 g were lost during the simulation. We give a more detailed discussion about the relevant timescales of the results in Sec. \ref{sec:timescl}

\section{Discussion}
\label{sec:disc}
\subsection{Validity of the results: relevant timescales}
\label{sec:timescl}
\paragraph{Coagulation timescale}
We use such a collision model in Sec. \ref{sec:res1} where we start the simulation with centimeter-sized particles and we follow how they grind down to smaller particles. The question arises, in what timescale can these particles form by coagulation? Is it valid to assume that these particles are present in the disk already after 100 orbits of the secondary? In order to answer this question, we estimate the coagulation timescale:
\begin{equation}
t_{\mathrm{coag}} = \frac{1}{n_d \Delta v \sigma_{cr}},
\end{equation}
where $n_d$ is the number density of the dust aggregates with a given size ($n_d=\rho_g/100/m$, where $\rho_g$ is the gas density, $m$ is the particle mass and we assumed a 100:1 gas to dust ratio.). Using typical density values at 1 AU in our models, assuming a particle mass of 1 g, and relative velocity of 10 cm/s, we get that $t_{\mathrm{coag}}=3\times10^3$ yrs (50 orbits). Thus we conclude that particles can reach the initial particle sizes that we use in Sec.\ref{sec:res1} in 100 orbits, of course under ideal conditions.

\paragraph{Viscous timescale of the gas}
The viscous timescale of the disk at a given distance is given by
\begin{equation}
t_{\mathrm{vis}}=r^2/ \nu,
\end{equation}
where $r$ is the distance from the central star, $\nu$ is the viscosity which is defined in Eq. \ref{eq:nuT}. The viscous timescale in our model at 1 AU distance is of the order of $10^4$ yrs using $\alpha = 10^{-2}$, and $10^5$ yrs assuming $\alpha = 10^{-3}$. This corresponds to 150 to 1500 secondary orbits depending on $\alpha$. The viscous spreading and accretion of the disk restructures the density profile of the disk during one viscous timescale and our results have to be treated with caution if used for a longer timescale than this.

\paragraph{Radial drift timescale of the dust}
Radial drift ($v_D$) has two sources: the drift of individual particles with respect to the gas ($v_{d}$) and drift caused by accretion processes of the gas ($v_{dg}$), thus the total radial drift velocity is $v_D=v_{d}+v_{dg}$. We measure the total radial drift speed of particles with different Stokes numbers (or dimensionless stopping times) at 1 AU in the `single' simulation. The radial drift speed measured in that model is not affected by the eccentricity variations of the gas disk. The radial drift timescale is then the time the particle needs to drift into the star, e.g., drift 1 AU distance. It turns out that $t_{\mathrm{drift}} = 2 \times 10^4$ yrs (300 orbits), if the stopping time $t_s=10^{-3}$ $T_p$, where $T_p$ is the orbital period of the particle; $t_{\mathrm{drift}} = 10^4$ yrs (150 orbits), if $t_s=10^{-2}$ $T_p$; and $t_{\mathrm{drift}} = 2 \times 10^3$ yrs (only 30 orbits!), if $t_s=10^{-1}$ $T_p$. The particles located at the inner part of the disk in our simulations are lost into the primary within 150 orbits. Although the radial drift is also a serious problem for single disks, this poses an even stronger constraint on planet formation in binary disks. As the disk in this binary system is small (the size limited by the separation of the system), there is no reservoir of dust mass available as in disks around single stars. Unless there is a mechanism to recycle the dust, or stop the drift of the dust before it gets accreted onto the star, the total dust mass of the disk is lost within several hundred orbits. Therefore, the available time to form planetesimals from the dust aggregates can be limited to a very narrow window.

\subsection{A critical look at the collision model}
\label{sec:critic}
The considered sources of relative velocity between the dust and the gas are the radial drift and the perturbations in the gas velocity field due to the secondary. In reality, however, particles are also stirred up by turbulent eddies, particles settle toward the midplane of the disk and these processes contribute significantly to the relative velocity between particles. Therefore, our models underestimate the true relative velocities by neglecting these effects, thus the final particle masses and stopping times are probably somewhat overestimated. 

We do not consider collisions between similar sized particles. Determining a realistic collision speed (e.g., relative velocity at distance = 0) of similar sized aggregates in hydrodynamical simulations is non-trivial. One has to subtract the Kepler velocity from the velocity vectors of the two particles to get rid of the artifact of the Kepler shear (see e.g., \cite{Lyra2009a}). Even after this, the relative velocity of the particles inside the same grid cell strongly depends on the distance between these aggregates. A true collision speed can only be determined by extrapolating the relative velocity to zero distances using sufficiently large number of particle pairs in a statistical way. We therefore only consider the relative velocity between tiny micron sized particles (that essentially move with the gas) and the dust. These relative velocities are defined at the exact same location, thus the velocities used throughout the paper are true collision speeds. 

Furthermore, we only consider two collision types: sweeping up monomers, and erosion by monomers. There are however nine different collision types defined in the collision model of \cite{Guttler2010}. The most critical of them is bouncing. As we only calculate collision speeds between a monomer and an aggregate, currently we are unable to consider any other collision type but the ones described above. However, collisions between similar sized bodies can be more destructive. Particles could fragment into smaller pieces, or the growth could halt due to bouncing \citep{Zsom2010}. 

Considering all these limitations, our results can only be regarded as upper estimates of the final mass and stopping time distribution of dust in binary systems.



\subsection{Possible planetesimal formation processes}
We investigated dust coagulation/fragmentation in this work and found that the maximum aggregate sized formed by coagulation is roughly 1 g in the inner disk and $10^{-4}$ g in the outer disk in binary systems. We discuss in this section the possible methods to form planetesimals from these aggregates. We consider favorable places for coagulation in the disk, and also particle concentration mechanisms, where planetesimals can directly form from dust aggregates.

\begin{figure*}
  \includegraphics[width=0.5\textwidth]{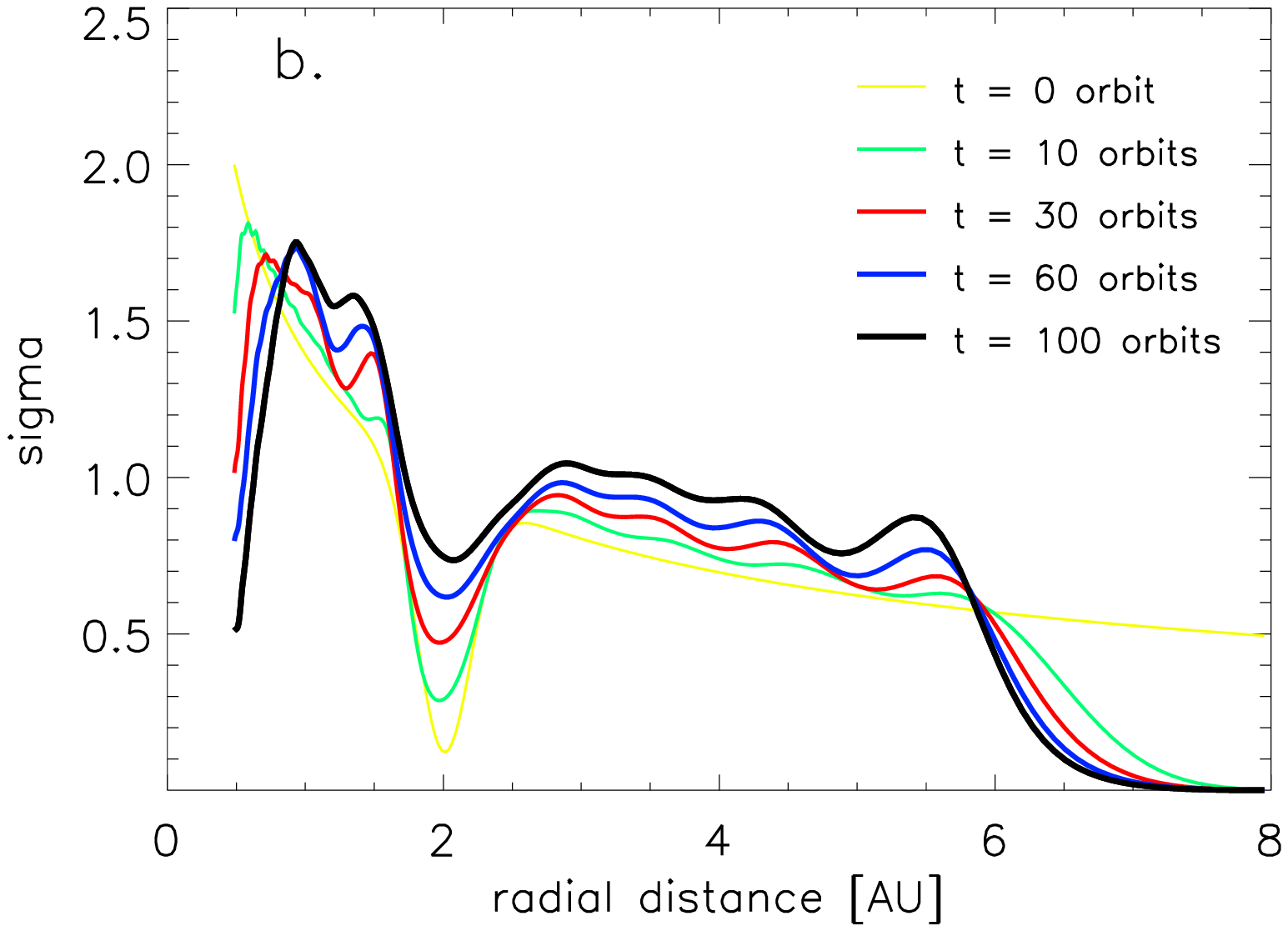}
  \includegraphics[width=0.5\textwidth]{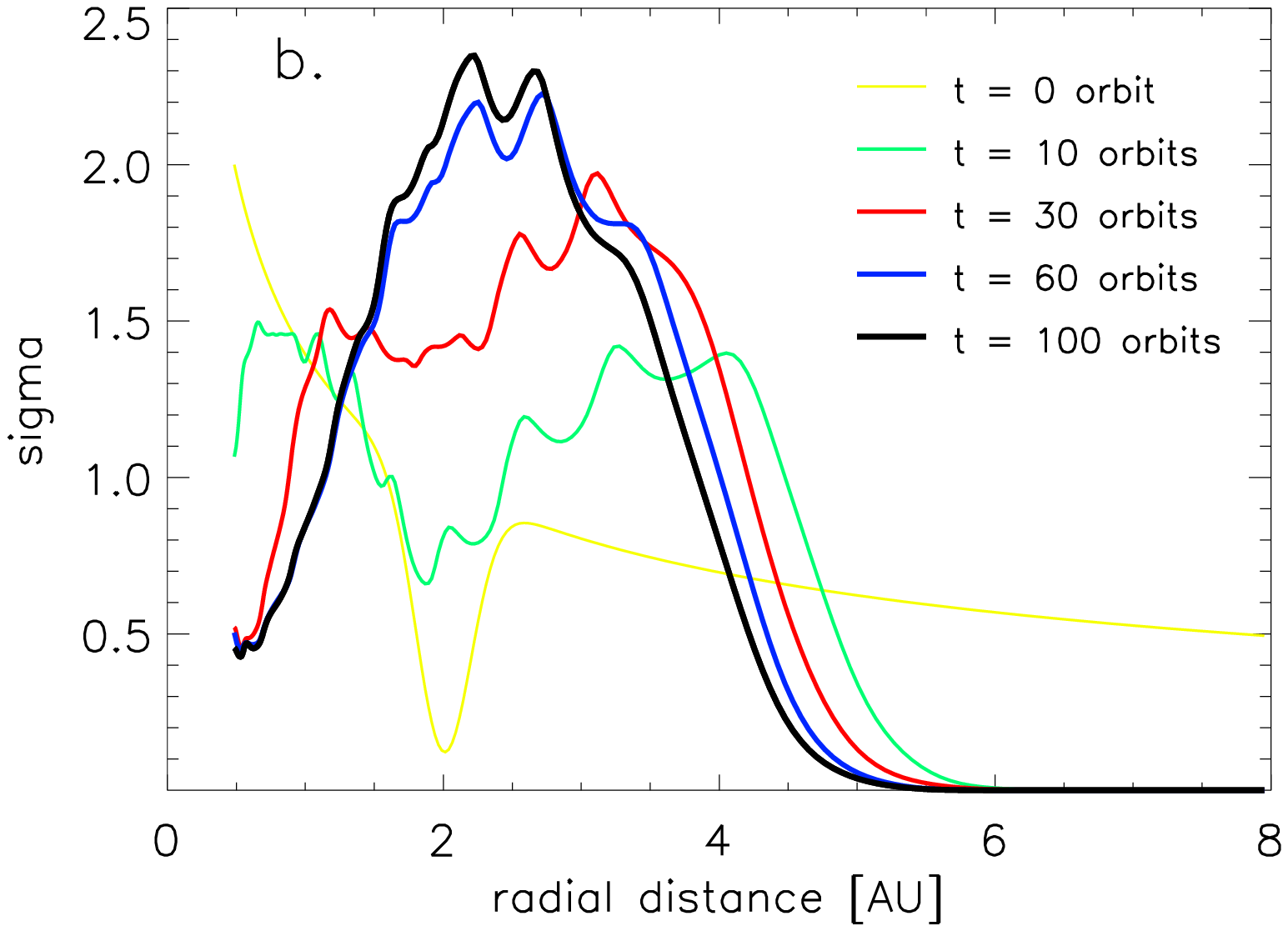}  
  \caption{The surface density profile of a disk with an artificial pressure bump at t=0, 10, 30, 60, and 100 orbits in case of a secondary on a circular orbit (a) and on an eccentric orbit with e=0.4 (b).}
  \label{fig:pbump}
\end{figure*}

\paragraph{Coagulation in a pressure bump.} \cite{Kretke2007} argued that a pressure bump forms around the snow-line, which stops the radial drift of the particles and concentrates them. Planetesimal formation can proceed in such locations as shown by the work of \cite{Brauer2008b}. If such a pressure bump exists in disks around primary stars, this is a favorable mechanism to form planetesimals. However, the recent work of \cite{Dzyurkevich2010} showed that the pressure bump may be milder than predicted by \cite{Kretke2007}. 

We performed two simulation with FARGO in which we simulated the stability of a simplified pressure bump in the presence of a circular and eccentric secondary. The initial surface density profile is altered to include a gauss-shaped `gap' (see \ref{fig:pbump}, black line at $t=0$ orbits). The gas velocity is altered in a way that the surface density profile is stable on a timescale shorter than the viscous timescale. In order to make the pressure bump stable on any timescale, the kinematic viscosity of the gas has to be radial dependent, which is currently not possible with FARGO. We choose the $\alpha$ parameter to be $\alpha = 10^{-5}$, thus the viscous timescale is $\sim 10^5$ secondary orbits. The azimuthal velocity on the right side of the gap is greater than the Kepler velocity, therefore this pressure bump would be strong enough to stop the radial drift of the particles, thus capture them there. We performed a test simulation where no secondary was present and the pressure bump did not change significantly over 150 orbits. 

Figure \ref{fig:pbump}a shows the evolution of the surface density in the presence of a secondary on a circular orbit. The secondary pushes the matter inward and excites waves in the gas. Therefore, the gap is gradually filled with gas on a timescale which is much shorter than the viscous timescale of the disk. This picture is even more dramatic in case the secondary is on an eccentric orbit with $e=0.4$ (see Fig. \ref{fig:pbump}b). The pressure bump disappears completely in 30 orbits. Despite of the simplicity of these simulations, it shows that the pressure bump can evolve on a shorter timescale than the viscous or coagulation timescale, namely on the orbital timescale of the secondary. If the pressure bump is not stable for even a coagulation timescale, planetesimals cannot be formed in such a pressure bump.

\paragraph{Other particle concentration mechanisms.} The other planetesimal formation mechanism was outlined by \cite{Johansen:2007p65}. This scenario assumes that a large amount of the solid material is presented in dm-sized boulders ($t_s \ge 0.1$) at the midplane of the disk. These boulders then concentrate in long-lived high pressure regions in the turbulent gas and initial over-densities are amplified further by the streaming instability. This mechanism forms 100 km sized objects on a very short timescale, however it is still debated whether this scenario works for smaller particles with $t_s < 10^{-2}$. 

\cite{Cuzzi2008} outlined an alternative concentration mechanism to obtain gravitationally unstable clumps of particles, which can then undergo sedimentation and form a `sandpile' planetesimal. In this model, turbulence causes dense concentrations of aerodynamically size-sorted, chondrule-sized particles. This mechanism effectively concentrates even $t_s=10^{-4}$ particles, therefore it seems a good candidate for planetesimal formation. However, this mechanism requires a critical amount of dust over-density in the clumps. The periodic perturbation from the secondary stirs up the dust particles, therefore it is necessary to examine whether particles can concentrate sufficiently in spite of the stirring of the secondary. 

\cite{Barge1995, Klahr1997} and recently \cite{Lyra:2008p625} showed that vortices accumulate dust particles very efficiently. Such vortices can be formed for example at the edge of the dead zone in disks. As discussed for the other concentration mechanism, it is necessary to examine whether the perturbation of the gas and the dust caused by the secondary does not prevent the concentration mechanism.



\section{Conclusions}
We have studied for the first time the coagulation/fragmentation processes of dust aggregates in a disk in binary systems. We have measured the relative velocity between the dust aggregates and the gas and assumed that this velocity is the same as the relative velocity between the dust and well-coupled monomers. Based on this, we have constructed a simple collision model. If the relative velocity is smaller than 1 m/s, the critical fragmentation velocity, the particle sweeps up monomers from the disk. If the relative velocity is greater than 1 m/s, the aggregate is eroded based on a recipe measured in laboratory experiments. Using this collision model, we have found that the eccentric secondary has two major impacts on the dust population in the disk. 
\begin{itemize}
\item A secondary on a circular orbit truncates the disk, therefore the particle masses/stopping times are decreased at the outer part of the disk. 
\item A secondary increases the eccentricity of the gas disk. If the eccentricity of the gas (therefore the deviation of the gas velocity field from the equilibrium velocity field in accretion disks) is increased, the relative velocity between the dust and the gas is greater. This in turn results in even smaller aggregate sizes throughout the disk.
\end{itemize}
Due to these effects, the average stopping time/mass of the particles in a disk with an eccentric secondary is reduced by one/four order/s of magnitude compared to the dust population which resides in a disk around a single star.

We have also studied the possibilities to form planetesimals from the dust aggregates investigates in this work (enhanced coagulation in a pressure bump and particle concentration mechanisms). We found that a pressure bump can evolve on shorter timescales than the viscous or coagulation timescale, namely on the orbital timescale of the companion. This effect, if it is confirmed by more detailed hydrodynamical studies, may suggest that a pressure bump in a binary system is not long lived enough for the formation of planetesimals. Particle concentration mechanisms in general need to cope with the continuous gravitational stirring and perturbations of the secondary. 

\begin{acknowledgement}
Andras Zsom would like to thank to Frederic Masset, Sijme-Jan Paardekooper, Wladimir Lyra, Anders Johansen, Chris Ormel, Til Birnstiel and Felipe Gerhard for useful discussions and help during the project. We also thank the anonymous referee for carefully reading the manuscript and helping us to improve it. A. Zsom acknowledges the support of the IMPRS for Astronomy \& Cosmic Physics at the University of Heidelberg.
\end{acknowledgement}

\end{document}